\documentclass[12pt,a4paper]{amsart}
\usepackage{amsmath,amssymb,epsfig,subfigure}

\newcommand{\e}{\epsilon}

\makeatletter \@addtoreset{equation}{section} \makeatother

\newtheorem{theorem}{Theorem}[section]

\newcommand{\ra}{\rightarrow}
\begin{document}
\title[Small dispersion limit of KdV]{
Numerical solution of the small dispersion limit of Korteweg de Vries 
and Whitham equations}
\begin{abstract}
The Cauchy problem for the Korteweg de Vries (KdV) equation with 
 small dispersion of order $\e^2$, $\e\ll 1$, 
is characterized by the appearance of a zone of rapid modulated oscillations
of wave-length of order $\e$.
These oscillations are approximately described by the elliptic solution of KdV where the amplitude,
 wave-number and frequency are not constant but evolve according to the Whitham equations.
In this manuscript we give a quantitative analysis of the 
discrepancy between
the numerical solution of the KdV equation in the small dispersion 
limit and  the corresponding approximate solution for values of 
$\epsilon$ between $10^{-1}$ and $10^{-3}$.
The numerical results are compatible with a difference of order $\e$ 
within the `interior' of the Whitham
 oscillatory zone, of order $\e^{\frac{1}{3}}$ at the left  boundary 
outside the Whitham zone and of order $\sqrt{\e}$ 
at the right boundary outside the Whitham zone.
\end{abstract}
\author[T. Grava]{T. Grava}
 \address{SISSA, via Beirut 2-4, 34014 Trieste, Italy} 
 \email{grava@fm.sissa.it}
\author[C. Klein]{C. Klein}
 \address{Max Planck Institute for Mathematics in the Sciences} 
 \email{klein@mis.mpg.de}
 \thanks{We thank S.~Bonazzola, B.~Dubrovin and J.~Frauendiener for helpful 
discussions and hints. CK and GT acknowledge support by the MISGAM program 
of the European Science Foundation. GT  acknowledges support by the RTN ENIGMA and  Italian COFIN 2004 ``Geometric methods in the theory of nonlinear waves and their applications''.}

\maketitle
\section{Introduction}
The purpose of this  manuscript  is the quantitative numerical  comparison of 
the solution of the Cauchy problem for the Korteweg de Vries equation    (KdV) 
\begin{equation}
\label{KdV}
u_t+6uu_x+\epsilon^{2}u_{xxx}=0,\quad u(x,0)=u_0(x),
\end{equation}
in the small dispersion limit  (small $\epsilon$) and the 
asymptotic formula obtained in the works of Lax and Levermore \cite{LL}, Venakides \cite{V2} and 
Deift, Venakides and Zhou \cite{DVZ} which describes the solution of the 
above Cauchy problem at the leading order as $\e\rightarrow 0$.
\begin{figure}[!htb]
\centering\epsfig{figure=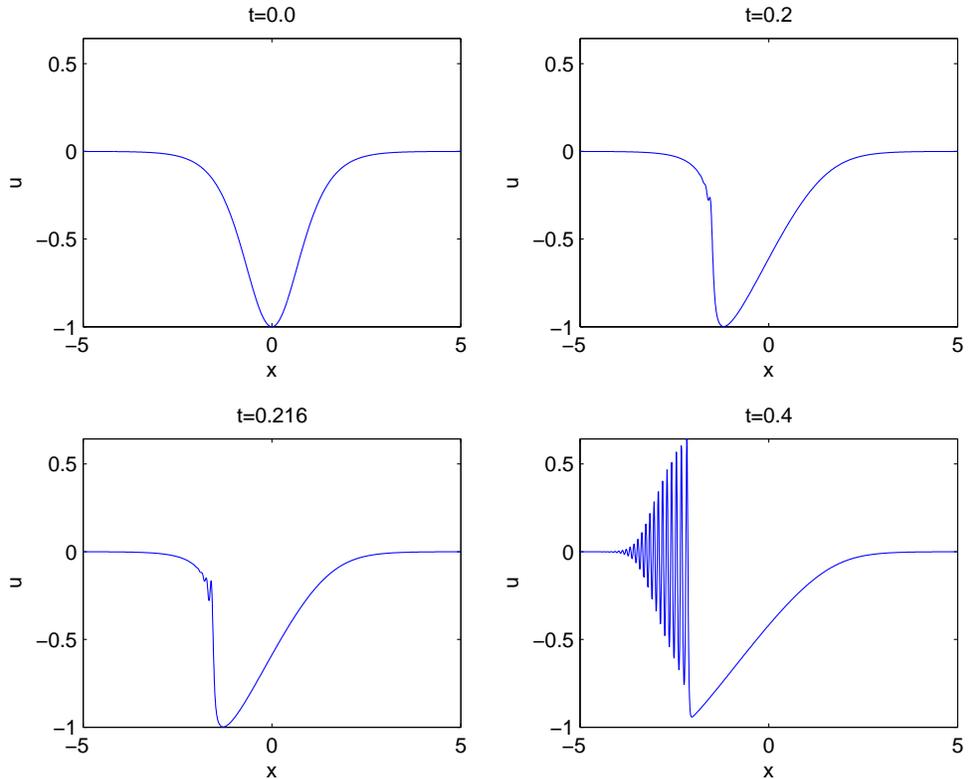, width=15.0cm}
\caption{The numerical solution of the KdV equation  at different times for 
the initial data 
$u_0(x)=-1/\cosh^2x$  and $\epsilon=10^{-1.5}$.}
\label{fig1}
\end{figure}
We study initial data with a negative hump and with a single minimum value at $x=0$ normalized 
to $-1$. The solution of the Cauchy problem
for the KdV equation is characterized by the appearance of a zone of 
fast oscillations of wave-length of order $\e$, see e.g.~Fig.~\ref{fig1}. 
These oscillations were called by Gurevich and Pitaevski
dispersive shock waves \cite{GP}.

\noindent
Following the work of \cite{LL}, \cite{V2} and  \cite{DVZ},
the rigorous theoretical  description of  the small dispersion limit of the 
KdV equation is   the following.
Let us define
\begin{equation}
\label{baru}
\bar{u}(x,t)=\lim_{\e\ra 0}u(x,t,\e).
\end{equation}
\noindent
1) for $0\leq t< t_c$, where $t_c$ is a critical time,  
the  solution $u(x,t,\epsilon)$ of the KdV  Cauchy problem  is approximated, 
for small $\e$, by the limit $\bar{u}(x,t)$ which solves the Hopf equation 
\begin{equation}
\label{Hopf}
u_t+6uu_x=0.
\end{equation}
Here $t_c$ is the time when the first
point  of gradient catastrophe appears in the solution 
\begin{equation}
\label{Hopfsol}
u(x,t)=u_0(\xi),\quad x=6tu_0(\xi)+\xi,
\end{equation}
of the Hopf equation. 
From the above, the time $t_c$ of gradient catastrophe can be
evaluated from the relation
\[
t_c=\min_{\xi\in\mathbb{R}}{\left[-\dfrac{1}{6u_0'(\xi)}\right]}.
\]
2) After the time of gradient catastrophe, 
the solution of the KdV equation is characterized by the
  appearance  of an interval of rapid modulated oscillations.  
According to the Lax-Levermore theory, the interval $[x^-(t), x^+(t)]$ of the oscillatory zone is 
independent of $\epsilon$. Here $x^-(t)$ and $x^+(t)$  are
 determined from the initial data and satisfy the condition  $x^-(t_c)=x^+(t_c)=x_c$ where $x_c$ is the $x$-coordinate of the point of gradient catastrophe of the Hopf solution.
Outside the interval  $[x^-(t), x^+(t)]$ the leading order asymptotics of $u(x,t,\e)$  as $\e\ra 0$  is described by the solution of the Hopf equation (\ref{Hopfsol}).
Inside  the interval  $[x^-(t), x^+(t)]$ the solution $u(x,t,\e)$  is  approximately  
described, for small $\e$,  by  the elliptic solution of KdV \cite{GP}, \cite{LL},\cite{V2},\cite{DVZ}
\begin{equation}
\label{elliptic}
u(x,t,\e)\simeq \bar{u}+ 2\e^2\frac{\partial^2}{\partial
x^2}\log\theta\left(\dfrac{\sqrt{\beta_1-\beta_3}}{2\e K(s)}[x-2 t(\beta_1+\beta_2+\beta_3) -q];\mathcal{T}\right)
\end{equation}
where now $\bar{u}=\bar{u}(x,t)$ takes the form 
\begin{equation}
\label{ubar}
\bar{u}=\beta_1+\beta_2+\beta_3+2\alpha,
\end{equation}
\begin{equation}
\label{alpha}
\alpha=-\beta_{1}+(\beta_{1}-\beta_{3})\frac{E(s)}{K(s)},\;\;\mathcal{T}=i\dfrac{K'(s)}{K(s)},
\;\; s^{2}=\frac{\beta_{2}-\beta_{3}}{\beta_{1}-\beta_{3}}
\end{equation}
with  $K(s)$ and $E(s)$ the complete elliptic integrals of the first 
and second kind, $K'(s)=K(\sqrt{1-s^{2}})$;
 $\theta$ is the Jacobi elliptic theta function defined by the 
Fourier series
\[
\theta(z;\mathcal{T})=\sum_{n\in\mathbb{Z}}e^{\pi i n^2\mathcal{T}+2\pi i nz}.
\]
For constant values of the  $\beta_i$ the formula (\ref{elliptic}) is an exact solution of KdV well
known in the theory of finite gap integration \cite{IM}, \cite{DN0}. On the contrary,  in  the description 
of the leading order asymptotics of $u(x,t,\e)$ as $\e\ra 0$,
 the quantities $\beta_i$ depend on $x$ and $t$  and evolve
 according to the Whitham equations \cite{W}
\[
\dfrac{\partial}{\partial t}\beta_i+v_i\dfrac{\partial}{\partial x}\beta_i=0,\quad i=1,2,3,
\]
where the speeds $v_i$ are given by the formula
\begin{equation}
    v_{i}=4\frac{\prod_{k\neq
     i}^{}(\beta_{i}-\beta_{k})}{\beta_{i}+\alpha}+2(\beta_1+\beta_{2}+\beta_{3}),
    \label{eq:la0}
\end{equation}
with $\alpha$ as in (\ref{alpha}). 
Lax and Levermore first derived, in the oscillatory zone, the expression (\ref{ubar}) for 
 $\bar{u}=\bar{u}(x,t)$ which clearly does not satisfy the Hopf equation. 
The theta function formula (\ref{elliptic}) for the leading order asymptotics 
of $u(x,t,\e)$ as $\e\ra 0$,  was derived in the work of Venakides and the phase  
$q=q(\beta_1,\beta_2,\beta_3)$  was derived in the work of Deift, Venakides and Zhou \cite{DVZ} for the case of pure radiation initial data, namely initial data with no solitons. 
However,  we verify numerically that their formula holds also for initial 
data with point spectrum.  
We give a formula for $q$ which looks different but which is equivalent to the one in \cite{DVZ}
\begin{equation}
\label{q0}
    q(\beta_{1},\beta_{2},\beta_{3}) = \frac{1}{2\sqrt{2}\pi}
    \int_{-1}^{1}\int_{-1}^{1}d\mu d\nu \frac {f_-( \frac{1+\mu}{2}(\frac{1+\nu}{2}\beta_{1}
	+\frac{1-\nu}{2}\beta_{2})+\frac{1-\mu}{2}\beta_{3})}{\sqrt{1-\mu}
    \sqrt{1-\nu^{2}}},
\end{equation}
where $f_-(y)$ is the inverse function of the decreasing part of the initial data.
The above formula holds till some time $T>t_c$.
For later times the formula has to be slightly modified (see  formula (\ref{qnm})).
The  function $q=q(\beta_1,\beta_2,\beta_3)$ is symmetric with respect to $\beta_1,\beta_2$ and 
$\beta_3$,  and satisfies a linear
over-determined system of Euler-Poisson-Darboux type. It  has been  
introduced in the work of Fei-Ran Tian \cite{FRT1} to study the 
Cauchy problem for the Whitham equations after the first breaking of the    Hopf solution. 
A formula for the phase in the multi-bump case is derived in \cite{EKV}.
 
\noindent
3) Fei-Ran Tian proved  that the description  in 1) and 2) is generic 
for some time after  the time  $t_c$ of gradient catastrophe \cite{FRT1}.

\begin{figure}[!htb]
\centering
\epsfig{figure=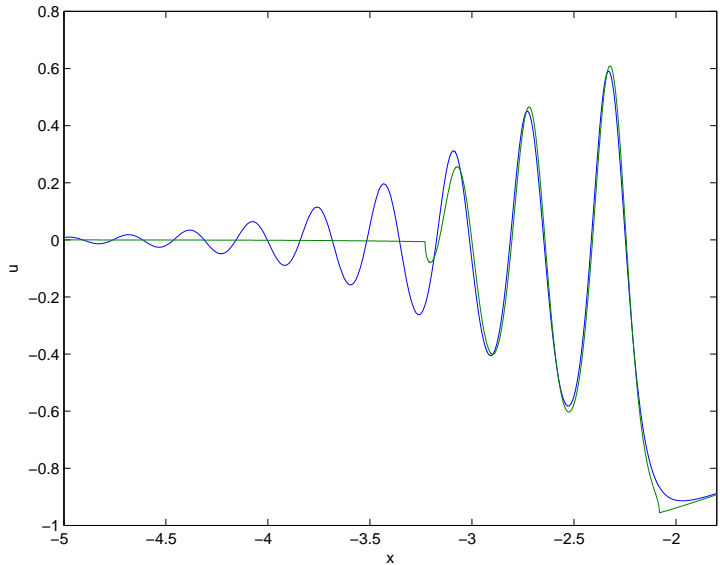, width=.7\textwidth}
\caption{The blue line shows the solution to the KdV equation for 
$\epsilon=0.1$ for the  initial data
$u_0(x)=-1/\cosh^2x$ at the time $t=0.4$, the green line the 
asymptotic solutions (\ref{Hopfsol}) and (\ref{elliptic}).}
\label{fig2in1}
\end{figure}

\noindent
In this paper we  compare numerically, outside and  inside the oscillatory region, 
the asymptotic formulas (\ref{Hopfsol}) and (\ref{elliptic}) 
with the solution of the KdV equation, see Fig.~\ref{fig2in1} for a 
plot of the two solutions 
for $\epsilon=0.1$.
Our main observations  are the following.
\begin{enumerate}
\item The oscillations of the numerical solution of the  KdV equation start at a time $t<t_c$.
Indeed  for the initial data $u_0(x)=-1/\cosh^2x$ the critical time  of the Hopf solution is given by 
$t_c=\sqrt{3}/8\simeq 0.2165$ and
 it can bee seen from Fig.~\ref{fig2} 
that the KdV solution has already  developed two oscillations at this time.
\begin{figure}[!htb]
\epsfig{figure=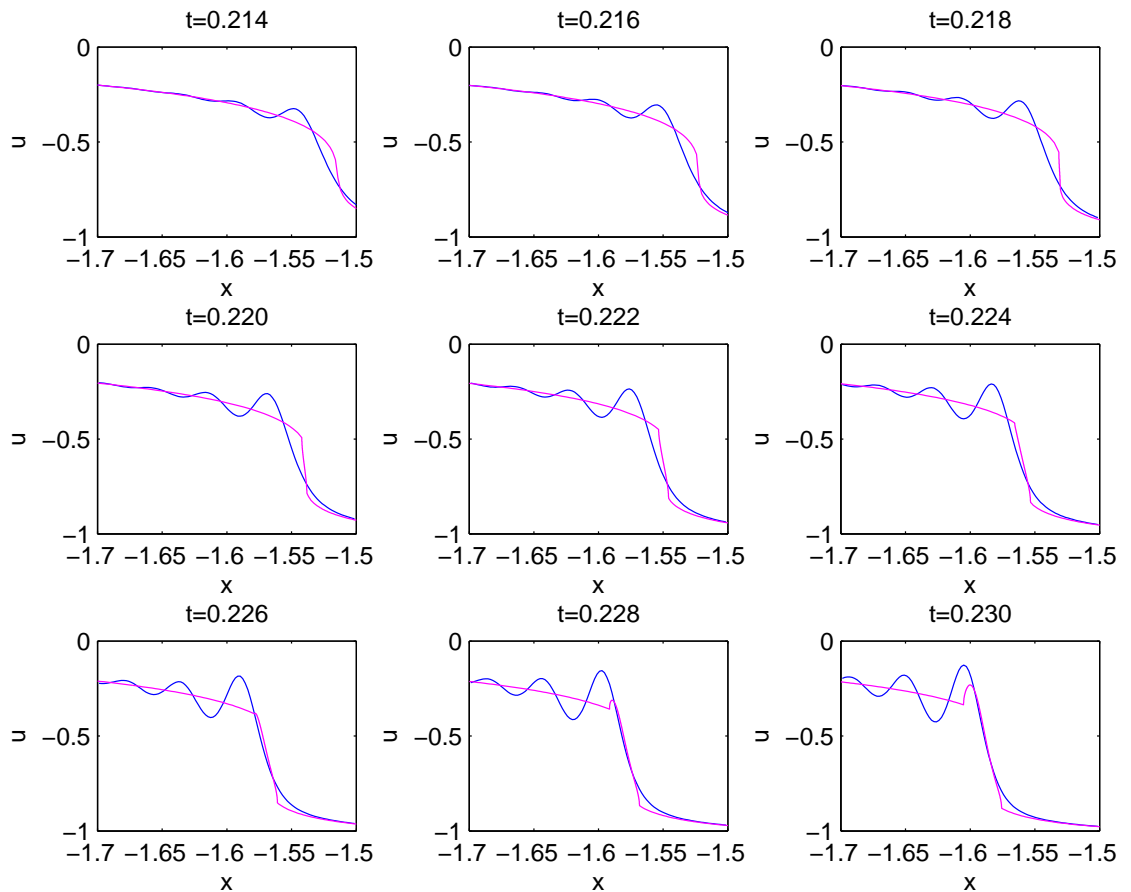, width=1.1\textwidth}
\caption{The blue line is the solution of the KdV equation for the 
initial data $u_0(x)=-1/\cosh^2x$ and $\epsilon=10^{-2}$, 
and the purple line is the corresponding  leading order 
asymptotics given by formulas  (\ref{Hopfsol}) and (\ref{elliptic}).
The plots are given for   different  times near the point of gradient catastrophe 
$(x_c,t_c)$ of  the Hopf solution. Here  $x_c\simeq -1.524 $, $t_c\simeq 0.216$.}
\label{fig2}
\end{figure}
\item The oscillatory interval of the KdV solution is bigger than the oscillatory interval 
$[x^-(t),x^+(t)]$ described by the leading order asymptotics given by formula (\ref{elliptic}). 
From the numerical simulation it can be seen that the KdV oscillatory interval is shrinking, 
as $\e\ra 0$, to the oscillatory interval  $[x^-(t),x^+(t)]$ (see Fig.~\ref{fig3} and
Fig.~\ref{fig4}). Let us define $\Delta^{\pm}_{hopf}:=
\mp(x_{hopf}^{\pm}/x^{\pm}-1)$ where $x_{hopf}^{\pm}$ are the 
values of $x$ for which the absolute value of the 
difference between the KdV solution and the 
asymptotic solution (\ref{Hopfsol}) are smaller than some fixed value (we 
take it to be $10^{-4}$ here) for all $\mp x>x_{hopf}^{\pm}$.
Then we numerically obtain that $\Delta^-_{hopf}\propto\e^{0.76}$ 
with standard error $0.028$ for the exponent.  However, this result has to be taken with care 
in view of the low number of points and the arbitrariness in the 
definition of the zone.  The zone $\Delta^+_{hopf}$  
is clearly shrinking with $\e$ but
the dependence on $\e$ does not seem to be described by a power law.
\begin{figure}[!htb]
\centering\epsfig{figure=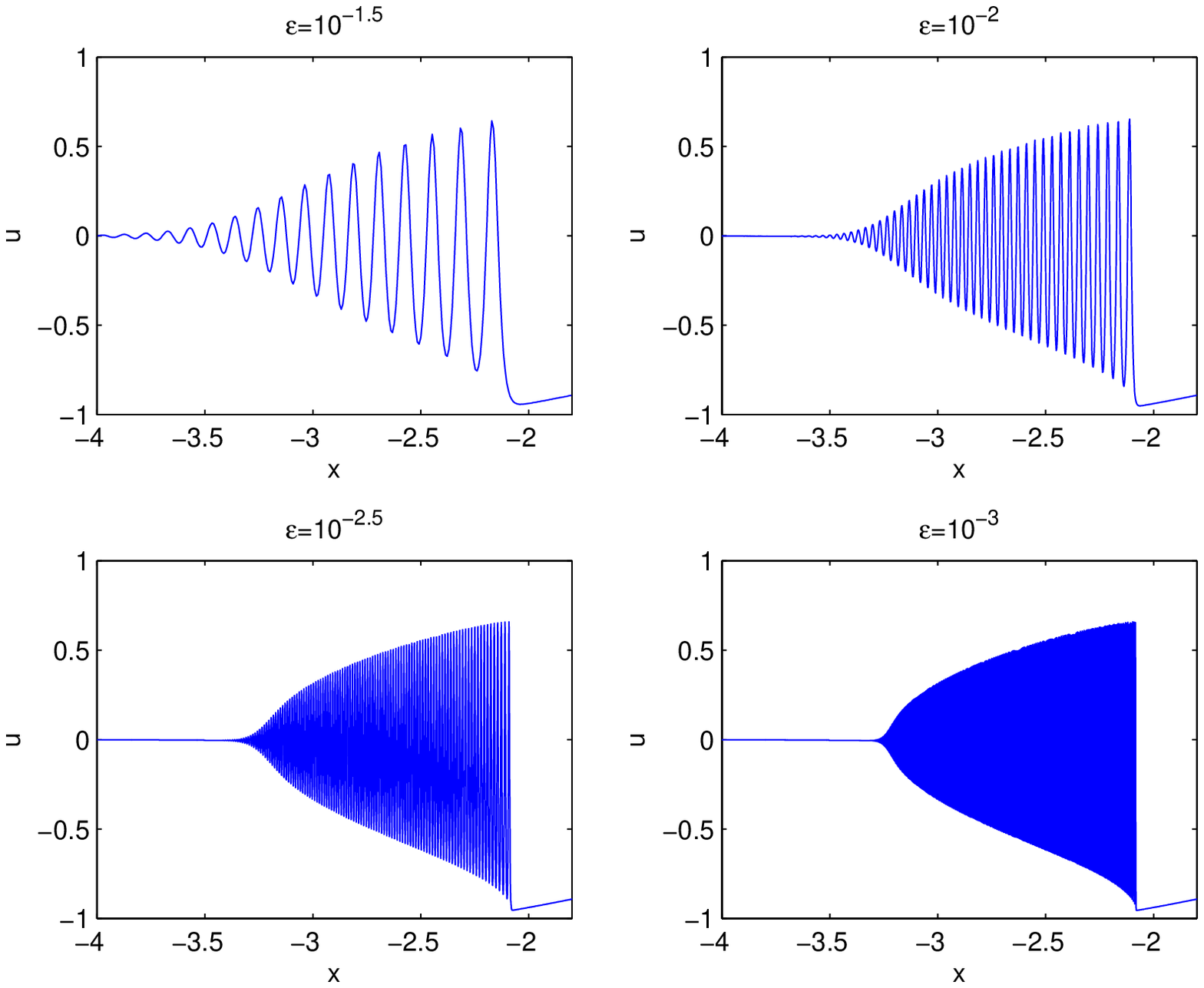, width=13.5cm}
\caption{The numerical solution of the KdV equation for the initial 
data $u_0(x)=-1/\cosh^2x$,  for  different values of $\epsilon$ and for fixed time $t=0.4$.}
\label{fig3}
\end{figure}
\item At the boundaries of the oscillatory region, the leading order asymptotics described by formula
(\ref{elliptic}) matches $C^0$, but not $C^1$, the solution of the Hopf equation 
(\ref{Hopfsol}) (see Fig.~\ref{fig2}).
We prove this statement analytically in Theorem~\ref{C0}.
\item The difference between the KdV solution and the asymptotic solution is decreasing with $\e$.
To define an error, we take the maximum 
of the absolute value of the difference between the solutions close 
to the center of the Whitham zone $[x^-(t),x^+(t)]$.
We find that this error is of order $\epsilon$ 
(more precisely $\e^{1.005}$ with standard error $0.05$).
At the left boundary of the oscillatory zone, and for $x<x^-(t)$  we obtain that 
the error is decreasing like  $\e^{0.35}$ with standard error $0.025$ 
which is roughly $\e^{\frac{1}{3}}$. 
The error for $x>x^+(t)$ is decreasing like $\e^{.525}$ with standard 
error $0.017$, which is roughly $\sqrt{\e}$.
\begin{figure}[!htb]
\centering
\mbox{\subfigure[KdV solution]
{\epsfig{figure=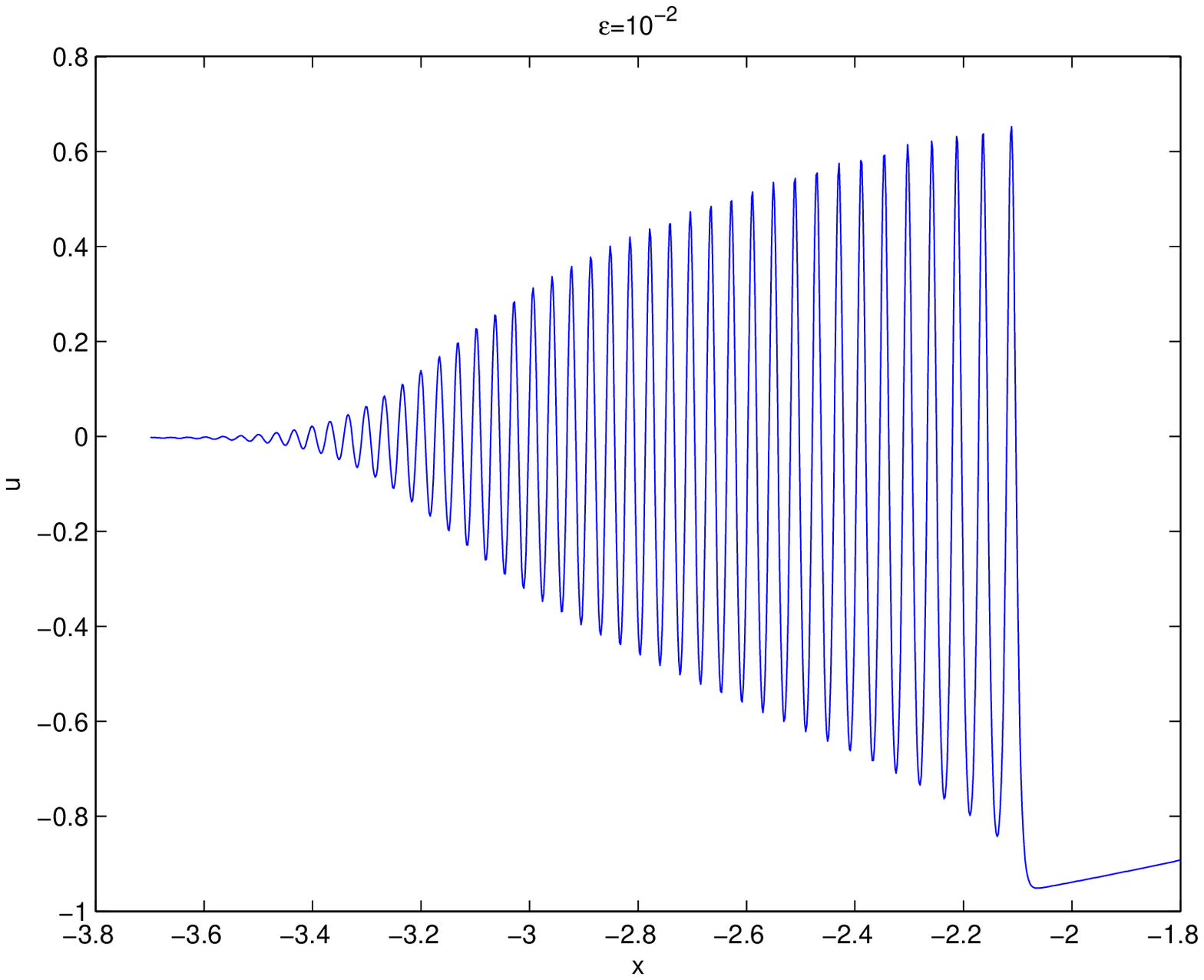, width=.5\textwidth}}\quad
\subfigure[Asymptotic solution]
{\epsfig{figure=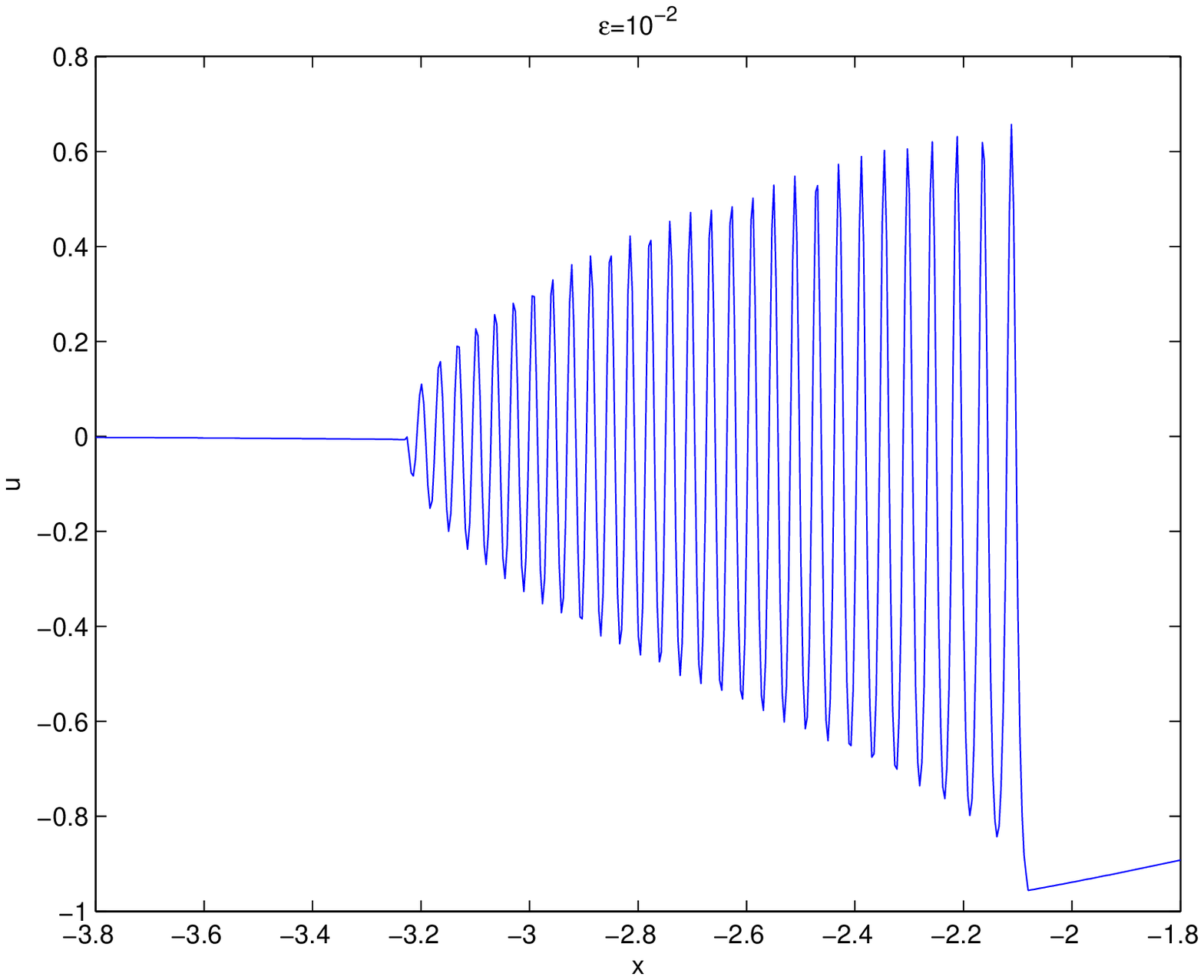, width=.5\textwidth}}}\newline
\caption{In  (a) the numerical solution of KdV is plotted for $t=0.4$ and  $\epsilon=10^{-2}$. In (b) the asymptotic formula (\ref{elliptic}) and (\ref{Hopfsol}) is plotted for the same values of $t$ and $\epsilon$.}
\label{fig4}
\end{figure}
\begin{figure}[!htb]
\centering
\epsfig{figure=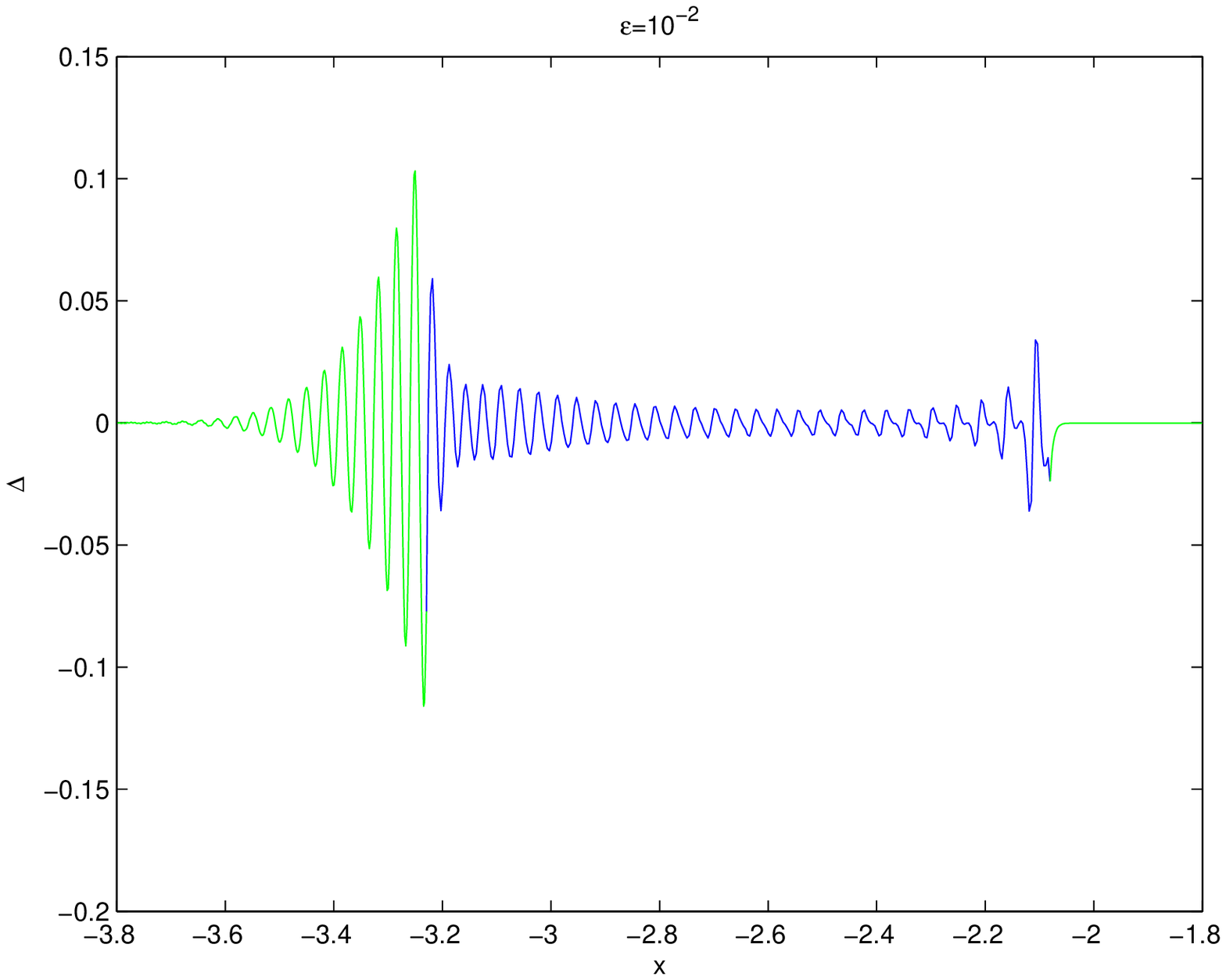, width=.6\textwidth}
\caption{The difference between the plots (a) and (b) of 
Fig.~\ref{fig3}; the Whitham zone is shown in blue, the exterior of 
this zone in green.}
\label{fig5}
\end{figure}
\end{enumerate} 
The manuscript is organized as follows: in the second section we give a brief analytical
description of the Lax-Levermore theory and the Whitham equations. In 
the third and fourth section
we give a description of the numerical algorithm used to solve the KdV equations and the 
Whitham equations, respectively. Readers not interested in the 
numerical details can skip this part. In the fifth section we present 
a detailed comparison of the numerical KdV solution in the limit of 
small dispersion and the corresponding asymptotic solution. In the 
last section we add some concluding remarks.
We concentrate on the explicit example of 
initial data $u_0(x)=-1/\cosh^2x$. The codes we developed are, 
however, suitable for any 
initial data with a single hump, which are rapidly decreasing at infinity.

\section{Lax Levermore theory and Whitham equations}
In this section we sketch the Lax-Levermore theory and its connection 
with the Whitham equations.
We consider initial data with one negative hump  that 
tends to zero fast enough as 
$x=\ra \pm \infty$. It is well known that the solution of the KdV equation 
can be obtained by the  inverse scattering method. 
Through this explicit expression  of  the KdV solution, 
Lax and Levermore  \cite{LL} for positive hump initial data  and  Venakides \cite{V1}
for negative hump initial data, managed to find the limit of the solution of (\ref{KdV}) as $\e\ra 0$ 
via a variational problem.
The KdV zero-dispersion limit exists in the distribution sense  and can be determined as follows:
\begin{equation}
\label{zerodispersion}
\bar{u}(x,t)=d-\lim_{\e\ra 0}u(x,t,\e)=2\partial^2_x\mathcal{Q}(x,t)-1,
\end{equation}
with
\begin{equation}
\label{variational}
\mathcal{Q}(x,t)=\inf_{\psi\in \mathcal{S}}\mathcal{Q}(\psi,x,t)
\end{equation}
where $\mathcal{S}$ is the set of all Lebesgue measurable function 
with support in 
 $[\min u_0(x),0]$. For our numerical 
purposes we do not use the Lax-Levermore-Venakides functional $\mathcal{Q}(\psi,x,t)$ 
and for this reason 
we omit the explicit formula but recall only  its properties.  
For each fixed $(x,t)$ the functional $\mathcal{Q}(\psi,x,t)$ assumes 
its minimum at exactly one element of the set $\mathcal{S}$ 
denoted by $\psi^*(x,t)$, moreover $\bar{u}(x,t)$ can be expressed in
terms of the endpoints of the support of $\psi^*(x,t)$.

To describe the support of the minimizer $\psi^*(x,t)$, 
Lax and Levermore consider the motion of 
the initial curve where each point of the curve has a different speed.
At time $t=0$ the curve is
$u=u_0(x)$ and the support of the minimizer is $[u_0(x),0]$. 
At later times the curve will, in general, be given 
by a multivalued  function in the $(x,u)$ plane with an odd number of branches.
  For  $0\leq t\leq t_c$, where $t_c$ is the time of gradient
 catastrophe for the Hopf equation, the curve is given by 
the solution   $u(x,t)$ of the Hopf equation (\ref{Hopf}) 
 and the support of the lax-Levermore-Venakides minimizer is $[u(x,t),0]$. 
The limit $\bar{u}(x,t)$  takes the form
\[
\bar{u}(x,t)=u(x,t).
\] 
 Soon after the time of  gradient 
catastrophe,
 the evolving curve is given by the multivalued function parameterized by 
three branches $0>\beta_1(x,t)>\beta_2(x,t)>\beta_3(x,t)>-1$, 
and the support of the minimizer is 
$[\beta_3,\beta_2]\cup[\beta_1,0]$ (see Fig.~\ref{figwhithopft}). 
The limit $\bar{u}(x,t)$ is expressed by the formula
\[
\bar{u}(x,t)=\beta_1(x,t)+\beta_2(x,t)+\beta_3(x,t)+\alpha(x,t),
\]
where $\alpha$ is defined in (\ref{alpha}).
In this case $\bar{u}(x,t)$ does not satisfy the Hopf equation 
(\ref{Hopf}) but a new set of equations enter the game.
Indeed the  $\beta_i$ as functions of $x$ and $t$ satisfy the Whitham equations \cite{W}
\begin{equation}
\label{whitham}
\dfrac{\partial}{\partial t}\beta_i+v_i
\dfrac{\partial}{\partial x}\beta_i=0,\quad i=1,2,3,
\end{equation}
where
\begin{equation}
    v_{i}=4\frac{\prod_{k\neq i}(\beta_{i}-\beta_{k})}{\beta_{i}+\alpha}+2\sum_{k=1}^{3}\beta_{k}.
    \label{eq:la}
\end{equation}
The equations (\ref{whitham}) are also called one-phase Whitham equations 
to distinguish them from the $n$-phase Whitham equations \cite{FFM} which are
derived when the support of the minimizer consists of $2n+1$ intervals.
In terms of the quantities $\beta_1,\beta_2$ and $\beta_3$  the approximate solution of 
$u(x,t,\e)$ as $\e\ra 0$ is given by the formula
\cite{V2}, \cite{DVZ}
\[
u(x,t,\e)\simeq \beta_1+\beta_2+\beta_3+2
\alpha+2\e^2\frac{\partial^2}{\partial
x^2}\log\theta\left(\dfrac{\sqrt{\beta_1-\beta_3}}{2\e K(s)}[x-2 t(\beta_1+\beta_2+\beta_3) -\phi];\mathcal{T}\right),
\]
with $\mathcal{T}$ and $K(s)$ as in (\ref{alpha}).
The   phase $\phi=\phi(\beta_1,\beta_2,\beta_3)$  is 
determined by the Deift-Venakides-Zhou formula \cite{DVZ} as follows. Let us introduce
the  functions $\rho(\lambda)$ and $\tau(\lambda)$ 
which 
 are the semiclassical 
approximation of the  reflection and transmission  coefficients  for the Schr\"odinger operator 
$-\e^2\partial_{xx}-u_0(x)$,
\begin{equation}
\label{rhotau}
\rho(\lambda)=\sqrt{-\lambda}x_+(\lambda)+\int_{x_+(\lambda)}^{+\infty}[\sqrt{-\lambda}
-\sqrt{u_0(y)-\lambda}]dy,\quad \tau(\lambda)=
\displaystyle\int_{x_-(\lambda)}^{x_+(\lambda)}\sqrt{\lambda-u_0(y)}dy,
\end{equation}
where $x_{\pm}(\lambda)$ are the solutions
of  $u_0(x_{\pm}(\lambda))=\lambda$.
Then the phase $\phi$ takes the form
\begin{equation}
\label{phase0}
\phi=\dfrac{1}{\pi}
\int_{I_1\cup I_2}\dfrac{\rho(\lambda)d\lambda}{\sqrt{(\lambda-\beta_1)(\lambda-\beta_2)(\lambda-\beta_3)}}-\dfrac{1}{\pi}\int_{[0,\eta]\backslash\cup I_j}\dfrac{i\tau(\lambda)d\lambda}
{\sqrt{(\lambda-\beta_1)(\lambda-\beta_2)(\lambda-\beta_3)}},
\end{equation}
where $I_1=[\beta_1,0],\;I_2=[\beta_3,\beta_2]$ and $\eta=-1$ when $t<T$ while
$\eta=\beta_3$ when $t>T$ where $T$ is the time when $\beta_3$ first reaches the minimum value
 $\beta_3=-1$. 

The problem of determining the small dispersion limit of KdV is  
reduced to solving the Lax-Levermore-Venakides  variational
problem (\ref{variational}) or 
 the Whitham equations (\ref{whitham}). In  \cite{MS} 
McLaughlin and Strain obtain numerically the weak limit $\bar{u}$ 
by solving the Lax-Levermore-Venakides  variational problem.
 In this manuscript we follow the latter approach, and we solve
numerically the Whitham equations (\ref{whitham}), 
thus determining the support of  the Lax-Levermore minimizer. 
The numerical solution of the Whitham equations has been already implemented 
in  the  early works of 
Gurevich and Pitaevski \cite{GP} and  Avilov and
Novikov \cite{AN} for step-like and cubic initial data. 
In this manuscript we implement a code which is suitable to rapidly 
decreasing initial data with an arbitrary single hump. 
As a concrete example we study  the case of
the initial data $u_0(x)=-1/\cosh^2 x$ in detail.

\subsection{The Cauchy problem for the Whitham equations}
In this subsection we mainly follow the work of Fei-Ran Tian 
\cite{FRT1}, \cite{FRT3}.
The Whitham equations are a system of hyperbolic PDEs defined for  
$\beta_1>\beta_2>\beta_3.$
Using the properties of the elliptic integrals
\begin{equation}
\label{elliptictrail}
K(s)=\dfrac{\pi}{2}\left(1+\dfrac{s}{4}+\dfrac{9}{64}s^2+O(s^3)\right),\quad
E(s)=\dfrac{\pi}{2}\left(1-\dfrac{s}{4}-\dfrac{3}{64}s^2+O(s^3)\right),
\end{equation}
and
\begin{equation}
\label{ellipticlead}
E(s)\simeq 1+\dfrac{1}{4}(1-s)\left[\log\dfrac{16}{1-s}-1\right],
\quad K(s)\simeq \dfrac{1}{2}\log\dfrac{16}{1-s},\quad \mbox{as}\;\;s\ra 1,
\end{equation}
we find that the speeds $v_i$ defined in (\ref{eq:la}) 
have the following behavior

\noindent
1) at $\beta_2=\beta_1$
\begin{align*}
&v_1(\beta_1,\beta_1,\beta_3)=v_2(\beta_1,\beta_1,\beta_3)=4\beta_1+2\beta_3\\
&v_3(\beta_1,\beta_1,\beta_3)=6\beta_3;
\end{align*}
2)  at $\beta_2=\beta_3$
\begin{align*}
&v_1(\beta_1,\beta_3,\beta_3)=6\beta_1\\
&v_2(\beta_1,\beta_3,\beta_3)=v_3(\beta_1,\beta_3,\beta_3)=12\beta_1-6\beta_3;
\end{align*}
The initial value problem for the Whitham equations is to determine the
solution of (\ref{whitham}) from the following boundary conditions:

\noindent
a) {\it Leading   edge:} 
\begin{equation}
\label{leadboundary}
\begin{split}
&\beta_1=\mbox{ the Hopf solution (\ref{Hopf})}\\ 
&\beta_2=\beta_3,
\end{split}
\end{equation}
b) {\it Trailing  edge:}
\begin{equation}
\label{trailboundary}
\begin{split}
&\beta_2=\beta_1\\
&\beta_3=\mbox{ the Hopf solution (\ref{Hopf})}.
\end{split}
\end{equation}
The Whitham equations are a set of quasi-linear hyperbolic PDEs \cite{L} that can be integrated by
a generalization of the method of characteristics.
Dubrovin and Novikov \cite{DN} developed a  geometric-Hamiltonian theory of the Whitham equations (\ref{whitham}). Using this theory, Tsarev \cite{T} was able to integrate the equations through the   
so called hodograph transform, which   generalizes the method of characteristics, 
and which gives the solution  of (\ref{whitham}) in the implicit form 
\begin{equation}
\label{hodograph}
x=v_it+w_i,\quad i=1,2,3,
\end{equation}
where  the $v_i$  are defined in (\ref{eq:la}) and the $w_i=w_i(\beta_1,\beta_2,\beta_3)$  
are obtained from an algebro-geometric procedure \cite{K} by the formula \cite{FRT1}
\begin{equation}
    w_{i} =
    \frac{1}{2}\left(v_{i}-2\sum_{k=1}^{3}\beta_{k}\right)\frac{
    \partial q}{\partial\beta_{i}}+q,\quad i=1,2,3.
    \label{eq:w}
\end{equation}
The function $q$ is defined by
\begin{equation}
\label{q}
    q(\beta_{1},\beta_{2},\beta_{3}) = \frac{1}{2\sqrt{2}\pi}
    \int_{-1}^{1}\int_{-1}^{1}d\mu d\nu \frac {f_-( \frac{1+\mu}{2}(\frac{1+\nu}{2}\beta_{1}
	+\frac{1-\nu}{2}\beta_{2})+\frac{1-\mu}{2}\beta_{3})}{\sqrt{1-\mu}
    \sqrt{1-\nu^{2}}}.
\end{equation}
In the above formula  $f_{-}(y)$ is the inverse function of the decreasing part of the initial data $u_0(x)$.
The above formula for  $q(\beta_{1},\beta_{2},\beta_{3})$ 
is valid as long as $\beta_1>\beta_2>\beta_3>-1$. When $\beta_3$ reaches the 
minimum value $-1$  and passes over the negative hump, then it is necessary to take into account also the increasing part of 
the initial data $f_+$ in formula  (\ref{q}). We denote by $T$ this time.
For $t>T>t_c$ we introduce the variable  $X_{3}$ defined by  
$u_{0}(X_{3})=\beta_{3}$ which is  still  monotonous. For values of $X_{3}$ beyond the hump, namely $X_3>0$,  we have to substitute  (\ref{q})  by the formula \cite{FRT3}
\begin{equation}
\label{qnm}
q(\beta_1,\beta_2,\beta_3)=\frac{1}{2\pi}\int_{\beta_2}^{\beta_1} \dfrac{d\lambda
}{
\sqrt{(\beta_1-\lambda)(\lambda-\beta_2)(\lambda-\beta_3)}}
\left(\int_{\beta_{3}}^{-1}
    \frac{d\xi
    f_{+}(\xi)}{\sqrt{\lambda-\xi}}+\displaystyle\int_{-1}^{\lambda}
    \frac{d\xi f_{-}(\xi)}{\sqrt{\lambda-\xi}}\right).
\end{equation}
Clearly the $w_i$ are constructed in such a way that the matching conditions
(\ref{trailboundary}) and  (\ref{leadboundary})  are satisfied. Indeed 

\noindent
at the  {\it trailing  edge: $\beta_1=\beta_2$} 
\[
 w_1(\beta_1,\beta_1,\beta_3)= w_2(\beta_1,\beta_1,\beta_3),\quad 
w_3(\beta_1,\beta_1,\beta_3)= \left\{
\begin{aligned}
     & f_-(\beta_3),\;\mbox{for}\; 
t<T  \\
     & f_+(\beta_3),\;\mbox{for} \;t>T
\end{aligned}
\right.,
\]
and at the {\it  leading edge: $\beta_2=\beta_3$} 
\[
w_2(\beta_1,\beta_3,\beta_3)= w_3(\beta_1,\beta_3,\beta_3),\quad 
w_1(\beta_1,\beta_3,\beta_3)=f_-(\beta_1).
\]
At the leading and trailing edge the system (\ref{hodograph}) becomes degenerate. 
To avoid degeneracy we rewrite  the system (\ref{hodograph}) in the  equivalent form
\begin{equation}
\label{lead0}
\left\{
\begin{aligned}
&\dfrac{1}{(\beta_1-\beta_2)K(s)}[(v_1-v_2)t+w_1-w_2]=0\\
&v_3t+w_3=x\\
&\dfrac{1}{(\beta_2-\beta_3)}[(v_2-v_3)t+w_2-w_3]=0.
\end{aligned}
\right.
\end{equation}
In the limit $\beta_2=\beta_3$ the system (\ref{lead0})  reduces to the system \cite{FRT1}, \cite{GT}
\begin{equation}
\label{lead}
\left\{
\begin{aligned}
&6\beta_1t+f_-(\beta_1)-x=0\\
&    \Phi(\beta_{3},\beta_{1})+6t =  0\\
    &\partial_{\beta_{3}}\Phi(\beta_{3},\beta_{1})  =  0
    \end{aligned}
\right.
\end{equation}
where  
\begin{equation}
    \Phi(\xi,\eta)= \frac{1}{2\sqrt{2}}\int_{-1}^{1}d\mu \frac{f'_-(
\frac{1+\mu}{2}\xi+\frac{1-\mu}{2}\eta)}{
    \sqrt{1-\mu}}=
\frac{1}{2\sqrt{\xi-\eta}}\int_{\eta}^{\xi}d\mu\frac{f'_-(\mu)}{
    \sqrt{\xi-\mu}}.
    \label{eq:Phi}
\end{equation}
The formula (\ref{lead}) has been obtained by using for the system (\ref{lead0})
the expansion (\ref{elliptictrail})
of the elliptic functions and the properties of the function  
$q=q(\beta_1,\beta_2,\beta_3)$  which is symmetric with respect to $\beta_1,\beta_2$ and $\beta_3$.
The solution of (\ref{lead}) yields $x$, $\beta_1$ and $\beta_3$ as a  function of $t$. 
In particular $x=x^-(t)$ defines the left boundary of the oscillatory  zone.
In order to get an approximation near $x_c$ of the function $x=x^-(t)$ we 
make a  Taylor expansion of the system (\ref{lead}) near 
$\beta_3=\beta_1=u_c$, $x=x_c$ and $t=t_c$. For this purpose  we first solve 
the last equation of (\ref{lead}) for $\beta_3=\beta_3(\beta_1)$ 
and enter the second equation of (\ref{lead}) with this solution.  
Using the Taylor expansion near
the point of gradient catastrophe $(x_c,t_c,u_c)$, we obtain
\begin{equation*}
\left\{
\begin{aligned}
&x-x_c\simeq 6(\beta_1-u_c)(t-t_c)+6u_c(t-t_c)+
\dfrac{1}{6}f'''_-(u_c)(\beta_1-u_c)^3\\
&6(t-t_c)+\dfrac{(\beta_1- u_c)^2}{2}\times\\
&\quad\left.\left(\dfrac{\partial^2}{\partial \beta_3^2}\Phi(\beta_3,\beta_1)\left(\dfrac{\partial\beta_3}{\partial\beta_1}\right)^2+2\dfrac{\partial^2 \Phi(\beta_3,\beta_1)}{\partial \beta_3\partial\beta_1}\dfrac{\partial\beta_3}{\partial\beta_1}+\dfrac{\partial^2}{\partial \beta_1^2}\Phi(\beta_3,\beta_1)\right)\right|_{\beta_1=\beta_3=u_c}\simeq   0
    \end{aligned}
\right.
\end{equation*}
Using the identities
\[
\left.\dfrac{\partial^2}{\partial \beta_3^2}\Phi(\beta_3,\beta_1)\right|_{\beta_1=\beta_3=u_c}=\dfrac{8}{15}f'''_-(u_c),\;\;\quad\left.\dfrac{\partial^2}{\partial \beta_3\beta_1}\Phi(\beta_3,\beta_1)\right|_{\beta_1=\beta_3=u_c}=\dfrac{2}{15}f'''_-(u_c),\;\;
\] 
and
\[
\left.\dfrac{\partial^2}{\partial \beta_1^2}\Phi(\beta_3,\beta_1)\right|_{\beta_1=\beta_3=u_c}=\dfrac{1}{5}f'''_-(u_c),
\]
we obtain
\begin{equation*}
\left\{
\begin{aligned}
&x-x_c\simeq 6(\beta_1-u_c)(t-t_c)+6u_c(t-t_c)+
\dfrac{1}{6}f'''_-(u_c)(\beta_1-u_c)^3\\
&6(t-t_c)+\dfrac{(\beta_1- u_c)^2}{12}f'''_-(u_c)\simeq   0.
    \end{aligned}
\right.
\end{equation*}
Substituting the value of $\beta_1-u_c$ from the second equation in the first, we finally obtain
\begin{equation}
\label{xm}
x^-_{app}(t)\simeq x_c+6u_c(t-t_c)-\dfrac{36\sqrt{2}}{\sqrt{f'''_-(u_c)}}(t-t_c)^{\frac{3}{2}},
\end{equation}
which is the semi-cubic law obtained in \cite{GP} for cubic initial data.

\begin{figure}[!htb]
\centering
\epsfig{figure=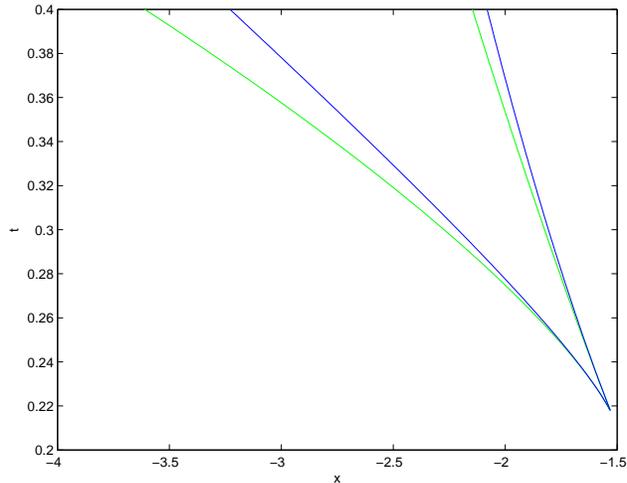, width=.6\textwidth}
\caption{The growth of the Whitham zone in the $(x,t)$ plane 
for the initial data $-1/\cosh^2 x$ (blue) and for cubic initial data 
(\ref{xm}) and (\ref{xp}) (green).}
\label{figwhitt}
\end{figure}

Using (\ref{ellipticlead}), the limit $\beta_2=\beta_1$  of the system (\ref{lead0})
 is  \cite{FRT1}, \cite{GT}
\begin{equation}
 \label{trail}
\left\{
\begin{aligned}
&    \Phi(\beta_{1},\beta_{3})+6t  = 0\\
&\int_{\beta_{3}}^{\beta_1}\sqrt{\lambda-\beta_3}[\Phi(\lambda,\beta_{3})+6t]d\lambda=0\\
&-x+6t\beta_3+f_-(\beta_3)=0,
\end{aligned}
\right.
\end{equation}
where the function $\Phi(\lambda,\beta_3)$ has been defined in (\ref{eq:Phi}).
The solution of (\ref{trail}) as a function of $t$ defines $x=x^+(t)$, which gives
the right boundary of the oscillatory zone. As for the leading edge, we can obtain
the approximate behavior of the function $x^+(t)$ near the point of gradient catastrophe 
(see Fig.~\ref{figwhitt})
\begin{equation}
\label{xp}
x^+_{app}(t)\simeq x_c+6u_c(t-t_c)+\dfrac{4\sqrt{10}}{3\sqrt{f'''_-(u_c)}}(t-t_c)^{\frac{3}{2}}.
\end{equation}

In the system (\ref{trail}), the  formula (\ref{eq:Phi}) 
for $\Phi(\beta_{1},\beta_{3})$   holds till the time $T>t_c$ when
$\beta_{3}=-1$ which can lead to $\beta_{3}$ being
non-monotonous. 
This time is given by the solution of the equations
\begin{equation}
\label{hump}
\left\{
\begin{split}
&x+6T=0\\
&\Phi(\beta_{1},-1)+6T  =  0\\
& \int_{-1}^{\beta_1}\sqrt{\lambda+1}[\Phi(\lambda,-1)+6T]d\lambda  =
0.
\end{split}
\right.
\end{equation}
For $t>T$ the function $\Phi(\lambda,\beta_3)$ appearing in the system 
(\ref{trail}) 
has to be modified to the form
\begin{equation}
    \Phi(\lambda,\beta_{3}) = \frac{1}{2\sqrt{\lambda-\beta_{3}}}
    \left(\int_{\beta_{3}}^{-1}
      \frac{dyf_{+}^{'}(y)}{\sqrt{\lambda-y}}+\int_{-1}^{\lambda}
      \frac{dyf_{-}^{'}(y)}{\sqrt{\lambda-y}}\right)
    \label{eq:Phi2}.
\end{equation}

\begin{figure}[!htb]
\centering
\epsfig{figure=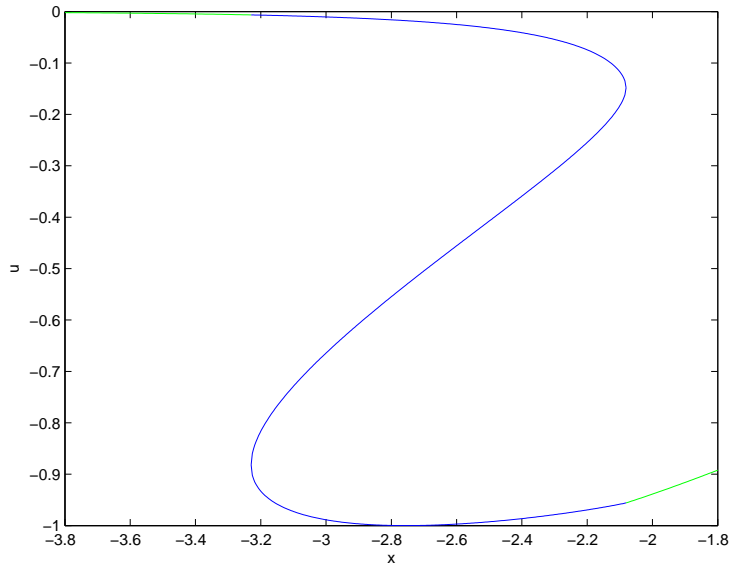, width=.7\textwidth}
\caption{ The blue line is the solution of the Whitham equations
$\beta_1>\beta_2>\beta_3$, the green line the solution $u(x,t)$ 
 of the Hopf equation (\ref{Hopf}) for the  initial data
$u_0(x)=-1/\cosh^2x$ and the time $t=0.4$.
The Whitham solution is attached  $C^1$ to the Hopf solution.}
\label{fig6b}
\end{figure}

The boundary conditions (\ref{leadboundary}) and (\ref{trailboundary})
 guarantee that the solution of the Whitham equations is attached in a $C^1$ way in the $(x,u)$ plane to the solution of the Hopf equation (see Fig~\ref{fig6b}).
Namely when $\beta_2=\beta_3$, the derivative 
 $\partial_x\beta_1$ is continuously attached  
to $\partial_x u(x,t)$ where $u(x,t)$ is the solution of the Hopf 
equation. The same holds in the case   $\beta_1=\beta_2$.
The partial derivatives $\partial_x \beta_i$, $i=1,2,3$ have been obtained 
in \cite{G2}  and take the form
\begin{equation}
\label{derivatives}
\partial_x \beta_i=\dfrac{\alpha+\beta_i}
{\prod_{j\neq i}(\beta_i-\beta_j)\partial_{\beta_i}Q},
\end{equation}
where $Q=Q(\beta_1,\beta_2,\beta_3)$ is given by
\[
Q(\beta_1,\beta_2,\beta_3)=\sum_{j=1}^3\partial_{\beta_j}q(\beta_1,\beta_2,\beta_3).
\]
To simplify the calculation we study $\partial_x\beta_1$ in 
the limit  $\beta_2\rightarrow \beta_3$ and  $\partial_x\beta_3$ 
in the limit  $\beta_2\rightarrow \beta_1$ 
before $\beta_3$ passes over the negative hump.  
Fixing $\beta_2=v+\delta$,  $\beta_3=v-\delta$, we obtain  for $\delta\rightarrow 0$, with 
the expansion (\ref{elliptictrail}), the relation
\begin{equation}
\label{beta1x}
\partial_x \beta_1=\dfrac{1}{6t+f'_-(\beta_1)}+O(\delta),
\end{equation}
Relation  (\ref{beta1x}) shows that, 
in the limit $\delta\ra 0$, the derivative $\partial_x \beta_1$
is converging to $\partial_x u(x,t)$ where $u(x,t)$ solves the Hopf equation.
Fixing  $\beta_2=v-\delta$, $\beta_1=v+\delta$ and evaluating 
$\partial_x\beta_3$  in the limit 
$\delta\rightarrow 0$, one obtains in the same way at the trailing edge 
\begin{equation}
\label{beta3x}
\partial_x \beta_3=\dfrac{1}{6t+f'_-(\beta_3)}+O(1/\log\delta),
\end{equation}
which shows that $\partial_x \beta_3\ra \partial_x u$
where $u$ solves the Hopf equation.
The solvability of the equations (\ref{hodograph}) and the systems
(\ref{lead}) and  (\ref{trail})  is guaranteed by the following theorem  \cite{FRT3}.
\begin{theorem}
Consider smooth initial data $u_0(x)$ with a single negative hump and suppose that $u_0(x)$ reaches its only minimum at $x=0$ where $u_0(0)=-1$. If the inverse function $f_-(u)$ of the decreasing part of $u_0(x)$ 
satisfies the conditions
\[
f''_-(u^*)=0,\quad f'''_-(u)<0,\;\;u\neq u^*
\]
where $u^*$ is the inflection point of $f_-(u)$, then the Whitham equation (\ref{whitham}) have a unique  solution
$\beta_1(x,t)>\beta_2(x,t)>\beta_3(x,t)$ for $t>t_c$ till  a short time after the time $T$.
\end{theorem}

In the following we show that the phase $\phi$ defined
in  (\ref{phase0}) can be expressed in terms of $q(\beta_1,\beta_2,\beta_3)$
defined in (\ref{q}) or (\ref{qnm}). Since this quantity is already 
numerically implemented in order to solve the Whitham equations, it 
is convenient to use this form of the phase.
\begin{theorem}
\label{C0}
The phase $\phi(\beta_1,\beta_2,\beta_3)$ defined in (\ref{phase0})
satisfies the following identity 
\begin{equation} 
\label{phi}
\phi(\beta_1,\beta_2,\beta_3)= q(\beta_1,\beta_2,\beta_3),
\end{equation} 
where 
$q=q(\beta_1,\beta_2,\beta_3)$ is  defined in  (\ref{q}) for $X_3<0$  and in 
(\ref{qnm}) for $X_3>0$, where $u_0(X_3)=\beta_3$.
\end{theorem}
\begin{proof}
Using the inverse functions $f_{\pm}$ we write $\rho(\lambda)$ and $\tau(\lambda)$ defined in (\ref{rhotau})
in the form
\[
\rho(\lambda)=\frac{1}{2}\int_{\lambda}^0\frac{f_+(\xi)}{\sqrt{\xi-\lambda}}d\xi,\quad
\tau(\lambda)=\dfrac{1}{2} \int_{-1}^{\lambda}\dfrac{f_+(\xi)d\xi}{\sqrt{\lambda-\xi}}-
\dfrac{1}{2} \int_{-1}^{\lambda}\dfrac{f'_-(\xi)d\xi}{\sqrt{\lambda-\xi}}.
\] 
Then the following identities hold
\begin{equation}
\begin{split}
&\int_{I_1\cup I_2}\dfrac{\rho(\lambda)d\lambda}{\sqrt{(\lambda-\beta_1)
(\lambda-\beta_2)(\lambda-\beta_3)}}=\\
&\quad\dfrac{1}{2}\int_{\beta_1}^0 f_+(\xi)d\xi\left(
\int_{\beta_3}^{\beta_2}+\int_{\beta_1}^{\xi}
\dfrac{d\lambda}
{\sqrt{(\lambda-\beta_1)(\lambda-\beta_2)(\lambda-\beta_3)(\xi-\lambda)}}\right)+\\
&\quad\dfrac{1}{2}\int_{\beta_3}^{\beta_2}\dfrac{d\lambda
}{\sqrt{(\lambda-\beta_1)(\lambda-\beta_2)(\lambda-\beta_3)}}
\displaystyle\int_{\lambda}^{\beta_1}\dfrac{f_+(\xi)}{\sqrt{\xi-\lambda}}
d\xi.
\end{split}
\end{equation}
Using the fact that the first term in the r.h.s. of the above relation is
identically zero and performing a change of coordinates of integration
in the second term, we obtain
\begin{align}
\nonumber
&\dfrac{1}{\pi}\int_{I_1\cup I_2}\dfrac{\rho(\lambda)d\lambda}{\sqrt{(\lambda-\beta_1)(\lambda-\beta_2)(\lambda-\beta_3)}}=\\
\label{tempq1}
&\frac{1}{2\sqrt{2}\pi}
    \int_{-1}^{1}\int_{-1}^{1}d\mu d\nu \dfrac{
f_+(\frac{1+\mu}{2}(\frac{1+\nu}{2}\beta_2
	+\frac{1-\nu}{2}\beta_3)+\frac{1-\mu}{2}\beta_1)}{ \sqrt{1-\mu}
    \sqrt{1-\nu^2} }.
\end{align}
For the term of $\phi$ containing the transmission coefficient $\tau$, 
we obtain  for $t<T$  
\begin{align}
\nonumber
&\dfrac{i}{\pi}\int_{[0,\eta]\backslash\cup I_j}\dfrac{\tau(\lambda)d\lambda}
{\sqrt{(\lambda-\beta_1)(\lambda-\beta_2)(\lambda-\beta_3)}}\\
&\nonumber
=\dfrac{i}{2\pi}\displaystyle\int_{\beta_2}^{\beta_1}\dfrac{d\lambda
}
{\sqrt{(\lambda-\beta_1)(\lambda-\beta_2)(\lambda-\beta_3)}}
\int^{\lambda}_{\beta_3}\dfrac{f_+(\xi)-f_-(\xi)}{\sqrt{\lambda-\xi}}d\xi\\
\label{tempq2}
&
=\frac{1}{2\sqrt{2}\pi}
    \int_{-1}^{1}\int_{-1}^{1}d\mu d\nu \dfrac{
f_+(\frac{1+\mu}{2}(\frac{1+\nu}{2}\beta_1
	+\frac{1-\nu}{2}\beta_2)+\frac{1-\mu}{2}\beta_3)}{ \sqrt{1-\mu}
    \sqrt{1-\nu^2} }\\
&\quad-\frac{1}{2\sqrt{2}\pi}
    \int_{-1}^{1}\int_{-1}^{1}d\mu d\nu \dfrac{
f_-(\frac{1+\mu}{2}(\frac{1+\nu}{2}\beta_1
	+\frac{1-\nu}{2}\beta_2)+\frac{1-\mu}{2}\beta_3)}{ \sqrt{1-\mu}
    \sqrt{1-\nu^2} }.
    \nonumber
\end{align}
Using the expression (\ref{tempq1}) and  (\ref{tempq2}) and 
the fact that the function $q(\beta_1,\beta_2,\beta_3)$ defined in (\ref{q}) 
is symmetric with respect to $\beta_1,\beta_2$ and $\beta_3$,
we derive the statement of the theorem for $t<T$.

\noindent
For $t>T$, repeating the same procedure above, we derive the relation
\begin{equation}
\label{tempq4}
\dfrac{i}{\pi}\int_{[0,\eta]\backslash\cup I_j}\dfrac{\tau(\lambda)d\lambda}
{\sqrt{(\lambda-\beta_1)(\lambda-\beta_2)(\lambda-\beta_3)}}=
\dfrac{1}{2\pi}\displaystyle\int_{\beta_2}^{\beta_1}\dfrac{d\lambda\displaystyle\int^{\lambda}_{-1}\dfrac{f_+(\xi)-f_-(\xi)}{\sqrt{\lambda-\xi}}d\xi}
{\sqrt{(\beta_1-\lambda)(\lambda-\beta_2)(\lambda-\beta_3)}}.
\end{equation}
Using the fact  that the expression (\ref{tempq1}) is symmetric  with respect to 
$\beta_1,\beta_2$ and $\beta_3$ we rewrite (\ref{tempq1}) in the form
\begin{equation}
\nonumber
\dfrac{1}{\pi}\int_{I_1\cup I_2}\dfrac{\rho(\lambda)d\lambda}{\sqrt{(\lambda-\beta_1)
(\lambda-\beta_2)(\lambda-\beta_3)}}=\dfrac{1}{2\pi}
\int_{\beta_2}^{\beta_1}\dfrac{d\lambda\displaystyle\int^{\lambda}_{\beta_3}
\dfrac{f_+(\xi)}{\sqrt{\lambda-\xi}}d\xi}{\sqrt{(\beta_1-\lambda)(\lambda-\beta_2)(\lambda-\beta_3)}}.
\end{equation}
Combining the above expression with (\ref{tempq4}) we arrive at the expression for 
the phase $\phi$ which coincides with  $q(\beta_1,\beta_2,\beta_3)$ 
defined in (\ref{qnm}).
\end{proof}

\noindent
In the following we show that the elliptic solution (\ref{elliptic}) 
attaches $C^0$ but not $C^1$ to the solution of the Hopf equation. 
This result is numerically obvious from Fig.~\ref{fig2} and Fig.~\ref{fig4}.
\begin{theorem}
The approximate solution of the Cauchy problem (\ref{KdV}) given by formula
(\ref{udn}) in the oscillatory zone and by the Hopf solution (\ref{Hopfsol}) outside the oscillatory zone is $C^0$ but not $C^1$ in the $(x,u)$ plane.
\end{theorem}
\begin{proof}
Let $u_{app}(x,t,\e)$ be the r.h.s. of (\ref{elliptic}) and rewrite formula (\ref{udn}) 
using the  Jacobi elliptic function $\mathrm{dn}$, where
\[
\mathrm{dn}(z\pi\theta^2(0;\mathcal{T}))=\sqrt{1-s^2}\dfrac{\theta(z;\mathcal{T})}{\theta(z+\frac{1}{2};\mathcal{T})}.
\]
The following identity holds \cite{Lawden}
\[
\dfrac{d^2}{dz^2}\log\theta(z;\mathcal{T})=4K^2(s)\left[\dfrac{1-s^2}{\mbox{dn}^2(2zK(s))}-\dfrac{E(s)}{K(s)}\right]
\]
so that, substituting the above into (\ref{elliptic}) we obtain
\begin{equation}
\label{udn}
u_{app}(x,t,\epsilon)=\beta_2+\beta_3-\beta_1+2\dfrac{\beta_1-\beta_2}{\mathrm{dn}^2\left(\dfrac{\Omega}{\epsilon}\right)},
\end{equation} 
where 
\begin{equation}
\label{omega}
\Omega=\sqrt{\beta_1-\beta_3}
(x-2(\beta_1+\beta_2+\beta_3)t-q).
\end{equation}
In the limit $\beta_2=\beta_3$ we have $\mathrm{dn}(z)\rightarrow 1$ so that 
\[
u_{app}(x,t,\epsilon)=\beta_1(x,t),
\] 
where $\beta_1(x,t)$ satisfies the Hopf equation because of the boundary condition (\ref{leadboundary}).

In the limit $\beta_2=\beta_1$, the function 
$\mathrm{dn}(z)\rightarrow \mathrm{sech} z$ and 
\begin{equation}
\label{omegatrail}
\Omega|_{[\beta_1=\beta_2]}=x-6t\beta_3-f_-(\beta_3)-[4t(\beta_1-\beta_3)-f_-(\beta_3)+\dfrac{1}{2}\int_{\beta_3}^{\beta_1}\dfrac{f_-(\xi)d\xi}{\sqrt{\beta_1-\mu}}]=0,
\end{equation}
because of (\ref{Hopfsol}) and (\ref{trail}), so that 
\[
u_{app}(x,t,\e)= \beta_3(x,t),
\]
where now $\beta_3$ satisfies the Hopf equation because of the boundary condition (\ref{trailboundary}).
Therefore $u_{appr}(x,t,\e)$  attaches $C^0$ to the solution of the Hopf equation in the limits
$\beta_2=\beta_1$ or $\beta_2=\beta_3$.

\noindent
Now we show that $u_{appr}(x,t,\e)$  is not attached $C^1$ to the solution of 
the Hopf equation.  
For this purpose we need to evaluate
$\partial_x u_{app}(x,t,\e)$, namely
\[
\partial_x u_{app}(x,t,\e)=\partial_x \beta_2+\partial_x \beta_3-\partial_x \beta_1+2\dfrac{\partial_x \beta_1-\partial_x \beta_2}{\mathrm{dn}^2(\Omega/\e)}
+4\dfrac{s^2(\beta_1-\beta_2)\mathrm{sn}(\Omega/\e) \mathrm{cn}(\Omega/\e)}{\mathrm{dn}^3(\Omega/\e)\e}\partial_x\Omega
\]
where $\Omega$ is defined in (\ref{omega}).
For simplifying the calculation, we compute 
the $x$-derivatives before $\beta_3$ reaches the minimum value $-1$. 
The calculation does not change in the general case, it is only more involved.
The derivatives $\partial_x\beta_i$ have been 
defined in (\ref{derivatives}). 
We first consider the leading edge, namely the case $\beta_2=\beta_3$.
Fixing $\beta_2=v+\delta$, 
$\beta_3=v-\delta$, we obtain  for $\delta\rightarrow 0$, with 
the expansion (\ref{elliptictrail}), the relation  (\ref{beta1x}) and 
\[
\partial_x \beta_2=\dfrac{1}{2\delta(v-\beta_1)\frac{\partial^2}{\partial v^2}\Phi(v,\beta_1)}+O(1),
\]
\[
\partial_x \beta_3=-\dfrac{1}{2\delta(v-\beta_1)\frac{\partial^2}{\partial v^2}\Phi(v,\beta_1)}+O(1).
\]
Furthermore as  $s\ra 0$,  $\mathrm{cn}(z)\rightarrow \cos(z)$, 
$\mathrm{sn}(z)\rightarrow \sin(z)$ and $\mathrm{dn}(z)\rightarrow 1$.
Combining the above limits and (\ref{elliptictrail}) we obtain that
\[
\partial_x u_{app}(x,t,\e)|_{[\beta_2=\beta_3]}=\infty,
\]
while  $\partial_xu(x,t)$, where $u(x,t)$ is the solution of the Hopf equation, remains finite at the leading edge.

At the trailing edge $\beta_2\rightarrow \beta_1$ or $s\rightarrow 1$.
In this limit $\mathrm{cn}(z)=\mathrm{dn}(z)=\mathrm{sech}(z)$ while $\mathrm{sn}(z)=\tanh(z)$ and the elliptic 
functions $E(s)$ and $K(s)$ behave as in (\ref{ellipticlead}).
We use the notation $\beta_2=v-\delta$, $\beta_1=v+\delta$ and evaluate 
the $x$-derivatives in the limit  $\delta\rightarrow 0$ obtaining  
\[
\partial_x \beta_1\simeq\dfrac{1/\partial_{v}\Phi(v,\beta_1)}{2\delta\log\dfrac{16(v-\beta_3)}{\delta} }
\]
\[
\partial_x \beta_2\simeq-\dfrac{1/\partial_{v}\Phi(v,\beta_1)}{2\delta\log\dfrac{16(v-\beta_3)}{\delta}}
\]
Using the above  relations, (\ref{beta3x})  and (\ref{omegatrail}) we find
\[
\partial_x 
u_{app}(x,t,\e)|_{[\beta_2=\beta_1]}= +\infty.
\]
while $\partial_xu(x,t)$ remains finite at the trailing edge.
\end{proof}

\subsection{The explicit example}
In this subsection we consider the explicit example of the initial data
\begin{equation}
    u_{0}(x)=-\frac{1}{\cosh^{2}(x)}
    \label{eq:u0}.
\end{equation}
For this initial data the point of gradient catastrophe can be 
evaluated analytically. It is given by 
\begin{equation}
    t_c=\dfrac{\sqrt{3}}{8},\;\;
    x_c=-\dfrac{\sqrt{3}}{2}+\log((\sqrt{3}-1)/\sqrt{2}),\;\;u_c=-2/3.
    \label{tcrit}
\end{equation}
The increasing and decreasing part  $f_{\pm}$ of the inverse function of $u_0(x)$ 
take the form
\begin{equation}
    f_{\pm}(y) = \ln \frac{1\pm\sqrt{1+y}}{\sqrt{-y}},
\quad 
    f_{\pm}'(y)=\mp\frac{1}{2y\sqrt{1+y}},
    \label{eq:fpm}
\end{equation}
where $-1\leq y<0$. 
Furthermore, we obtain an analytical expression for the function $\Phi(\lambda,\eta)$ defined in (\ref{eq:Phi})
\begin{equation}
\label{phi1}
\Phi(\lambda,\eta)=-\frac{1}{2\sqrt{|\lambda(\lambda-\eta)|}}
\mbox{arcsin}\sqrt{\left|\dfrac{\eta-\lambda}{\eta(\lambda+1)}\right|
}.
\end{equation}
Having the above explicit expressions, 
we can analytically solve the system (\ref{hump}), and obtain the time $T$ when $\beta_3=-1$,
\begin{equation}
\label{T}
T=\dfrac{\pi}{6\sqrt{3}},\;\quad x_T=-\pi/\sqrt{3},\quad\beta_1=-\dfrac{1}{4},\quad \beta_3=-1.
\end{equation}



\section{Numerical solution of the KdV equation}
In this paper we are interested in the numerical solution of the KdV 
equation for hump-like initial data which are rapidly decreasing for 
$|x|\to \infty$. Therefore it is legitimate to study the problem in a 
periodic setting of sufficiently large period. We use here a slightly 
modified version of Trefethen's code for the KdV equation 
(Chap.~10 in \cite{trefethen1}) which is available at 
\cite{trefethenweb}. 

The basic idea of the code is the use of a discrete Fourier 
transform in $x$ and  an integrating factor such that the time 
derivative contains the only linear term in $u$ in the equation. Let
$$
\hat{u}(t,k):= \int_{\mathbb{R}} u(t,x) e^{i k x }\, d x 
,
$$
be the Fourier transform of $u$ and let $\widehat{u^{2}}$ be 
the transform of $u^{2}$. Then the transformed KdV 
equation (\ref{KdV}) reads
\begin{equation}
    \hat{u}_{t}-\epsilon^{2}ik^{3}\hat{u}+3ik\widehat{u^{2}}=0
    \label{KdVfourier}.
\end{equation}
This is equivalent to
\begin{equation}
    \left(e^{-ik^{3}t}\hat{u}\right)_{t}+3ike^{-ik^{3}t}\widehat{u^{2}}= 0
    \label{KdVint}.
\end{equation}
The integrating factor avoids a stiff term in the equation
and thus allows for larger time steps. To solve equation (\ref{KdVint}) 
numerically we  use the \emph{Fast 
Fourier Transform} (FFT) in MATLAB for the $x$-dependence and a 
fourth-order Runge-Kutta method for the time integration. 

This code is perfectly adequate to solve the KdV equation for an $\epsilon$ 
of the order 1. However in the limit of small $\epsilon$ we are interested in 
here, the \emph{aliasing error} becomes significant. This error is due
to the pollution of the numerically calculated Fourier transform 
$\hat{u}$ by higher frequencies because of the truncation of the 
series, see \cite{canuto} for details. It
becomes important in dealing with nonlinearities in the equations. 
The error can be suppressed by putting a certain number of the high 
frequency components of $\hat{u}$ equal to zero after the nonlinear 
operations. As a rule of thumb it is sufficient to put roughly 
$1/3$ of the coefficients equal to zero \cite{canuto}. 
Thus effectively we are 
working with a lower resolution (2/3 of the number $N$ of modes 
given below), but this avoids high frequency noise and stabilizes the 
code. 

\begin{figure}[htb]
     \centering \epsfig{figure=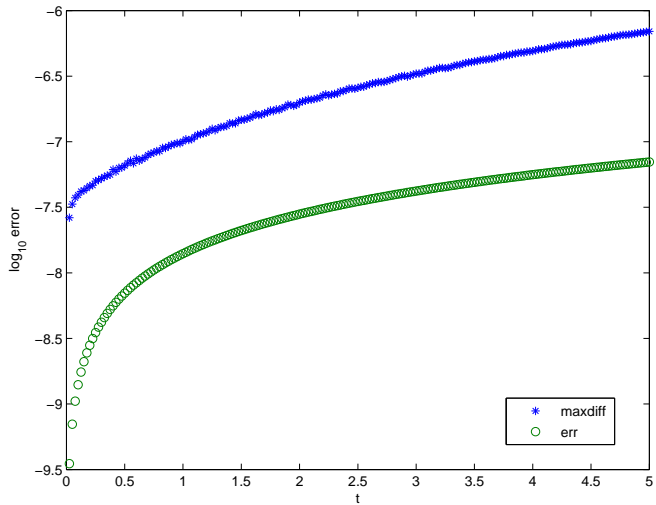, width=10.0cm}
    \caption{Numerical errors for the time evolution of 1-soliton 
    initial data: maximum of the absolute value of the difference between 
    exact and numerical solution (maxdiff) and deviation from energy 
    conservation (err).}
    \label{figerror}
\end{figure}

To test the accuracy of the code we consider an exact solution to the 
KdV equation, the 1-soliton $u=2/\cosh^{2}(x-x_{0}-4t)$ for 
$\epsilon=1$. The 
$x$-coordinate takes values in $[-\pi,\pi]L$ where the length $L$ is 
always chosen in a way that the coefficients for the initial data are 
or the order of the rounding error\footnote{MATLAB works internally
with a precision of $10^{-16}$. Due to rounding errors machine 
precision is generally limited to the order of $10^{-14}$.} 
for high frequencies. This 
reduces the error due to the discontinuity of the initial data at 
the interval boundaries to the order of the rounding error. We use 
the 1-soliton solution at time $t=0$ with $L=10$ and 
$x_{0}=-L$ as initial data and determine for each time step the 
difference of the exact and the numerical solution. The computation 
is carried out with $N=2^{11}$ 
modes (we will always use powers of 2 here since the FFT algorithm is 
most efficient in this case, but this is not necessary) and 4000 time 
steps for $t\in [0,5]$. The maximum of the absolute value of the difference 
between the numerical and the exact solution is shown as a function 
of time in Fig.~\ref{figerror}. 

An alternative test of the numerical precision is provided by 
conserved quantities as the energy,
\begin{equation}
    E = const \int_{-\infty}^{\infty}(2u^{3}-\epsilon^{2}u_{x}^{2})dx
    \label{energy}.
\end{equation}
This quantity is analytically conserved during time evolution, but 
numerically it will be a function of time due to unavoidable 
numerical errors. Since energy conservation is not implemented in  
the code, it provides a strong test for the numerical accuracy. We 
define the function $\mbox{err}$ via
\begin{equation}
    \mbox{err}:=1-E(t)/E(0)
    \label{errdef},
\end{equation}
where $E(t)$ is the numerically calculated energy which is obtained 
from the first component of the FFT of the integrand in (\ref{energy}) at each 
time step. For the example of the 1-soliton solution, this function 
is shown in Fig.~\ref{figerror}. It can be seen that the error 
obtained via the integral quantity is typically an order of magnitude 
higher than the maximal local difference of the exact and the numerical 
solution. In cases where no exact solution is known, we will use 
energy conservation as an indicator of the precision of the numerical 
solution. The number of modes and the time step will be chosen in a 
way that the value of the function $\mbox{err}$ is at least an order 
of magnitude lower than the precision of the numerical solution we 
are aiming at. 

\begin{figure}[htb]
     \centering \epsfig{figure=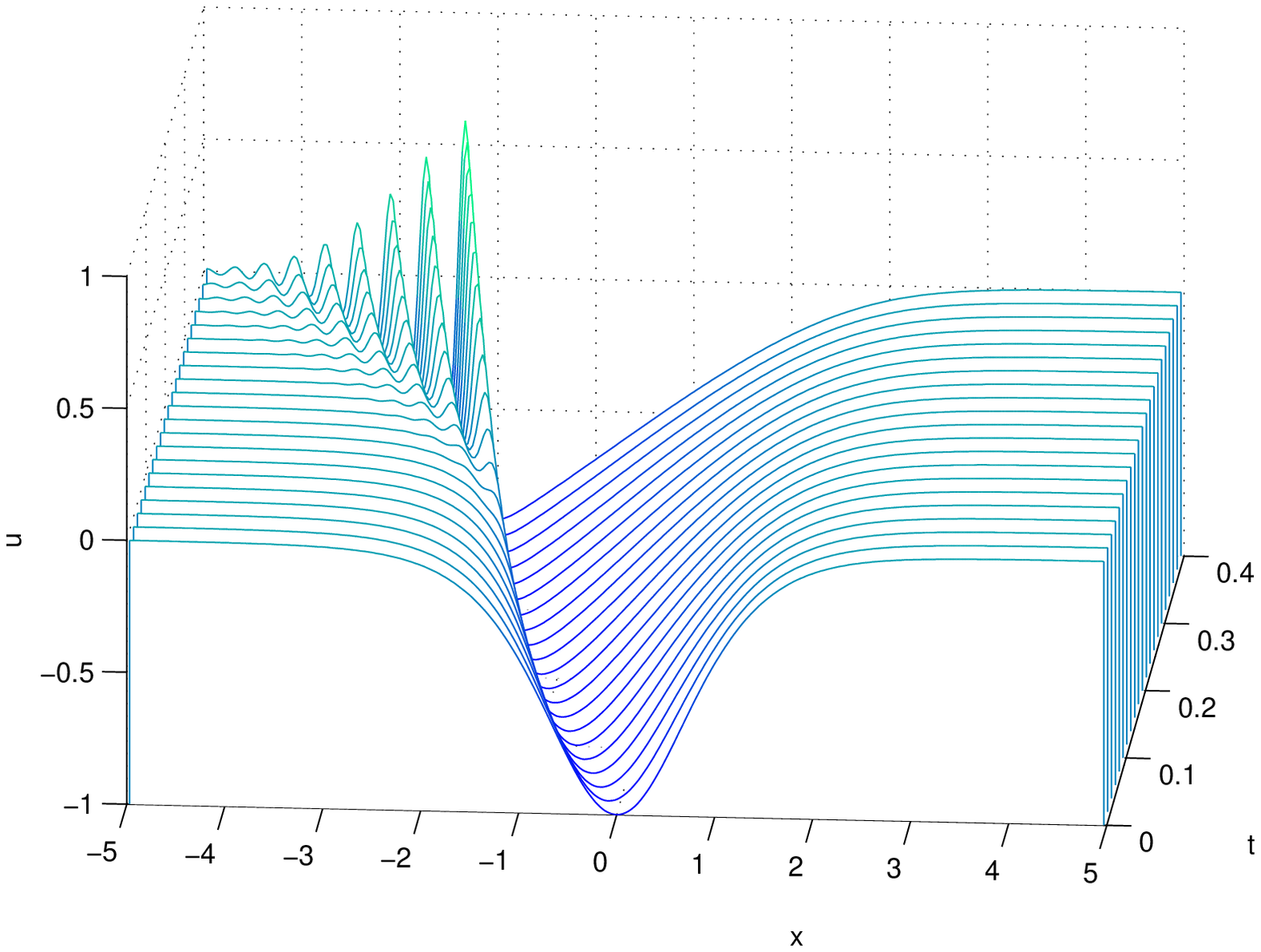, width=10.0cm}
    \caption{Numerical solution of the KdV equation with 
    $\epsilon=0.1$ and 
    initial data $-1/\cosh^{2}(x)$.}
    \label{figkdvp}
\end{figure}

\begin{figure}[htb]
    \centering 
     \epsfig{figure=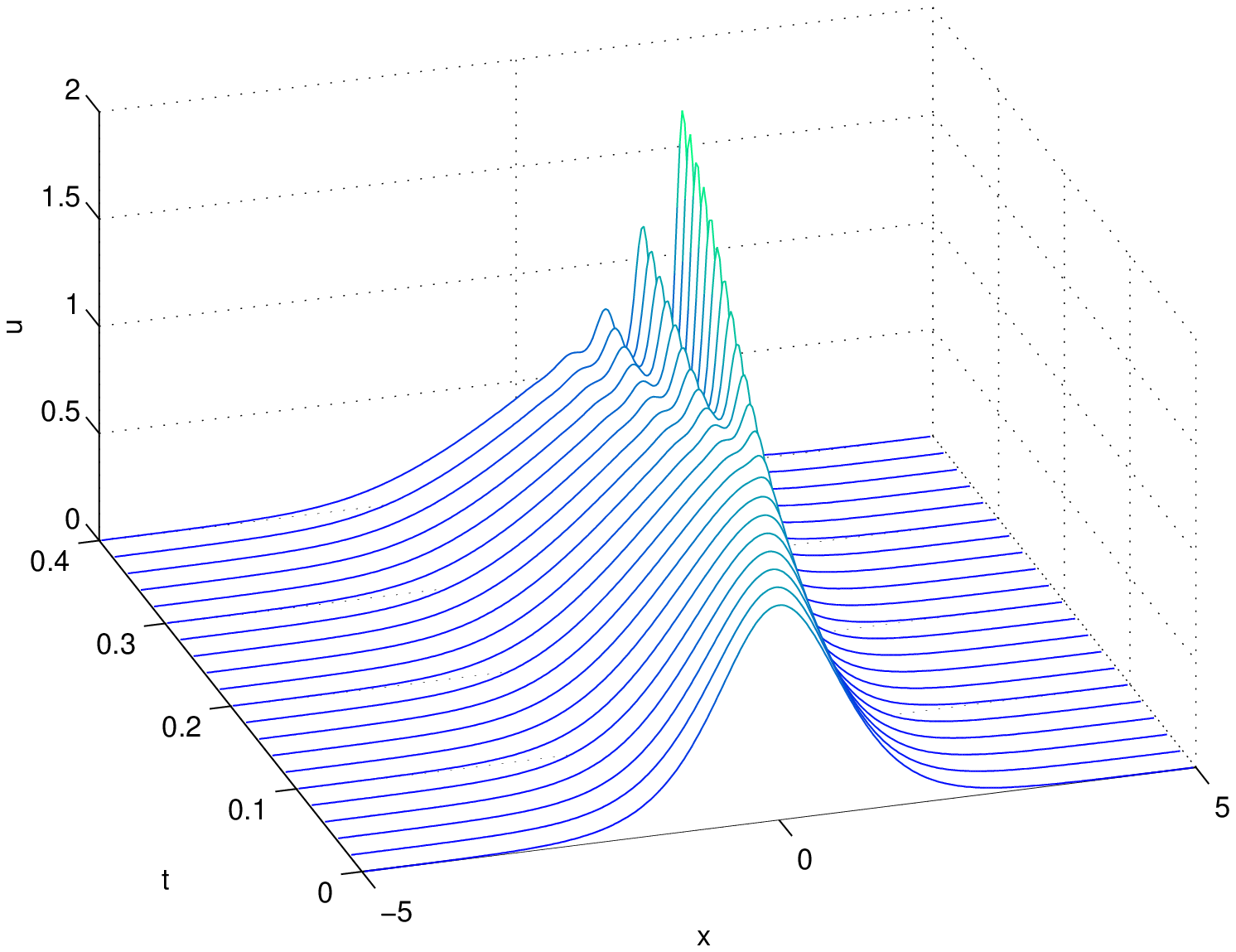, width=10.0cm}
    \caption{Numerical solution of the KdV equation with 
    $\epsilon=0.1$ and 
    initial data $1/\cosh^{2}(x)$.}
    \label{figkdvm}
\end{figure}

\begin{figure}[htb]
     \centering 
      \epsfig{figure=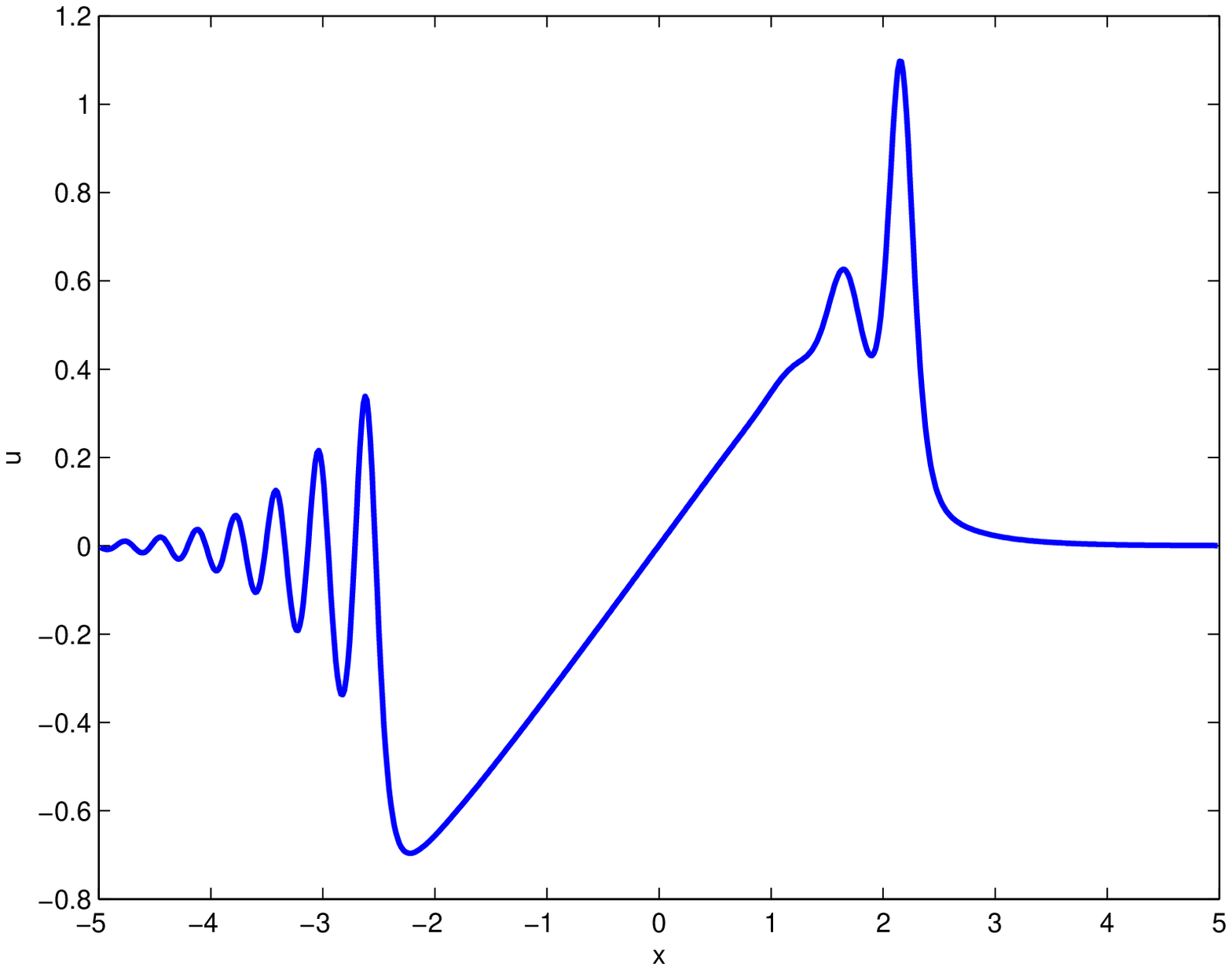, width=10.0cm}
    \caption{Numerical solution of the KdV equation with 
    $\epsilon=0.1$ and 
    initial data $2\sinh(x)/\cosh^{3}(x)$ for $t=0.4$.}
    \label{figkdvder}
\end{figure}

To study the small dispersion limit of KdV solutions, we consider 
hump-like data of the shape of the 1-soliton for both signs,
$u=\pm 1/\cosh^{2}(x)$. We show plots for the evolution of negative 
initial data in Fig.~\ref{figkdvp} 
and for positive initial data in 
Fig.~\ref{figkdvm} for $\epsilon=0.1$ and $t\in [0,0.4]$. 
To show that the form of the oscillations for positive and negative 
initial data is typical, we consider the evolution of the initial data 
$u=2\sinh(x)/\cosh^{3}(x)$ at time $t=0.4$ for $\epsilon=0.1$ 
in Fig.~\ref{figkdvder}. 

For a given value of $\epsilon$ we need a spatial resolution $2\pi 
L/N$ of at least $\epsilon$. We generally try to be an order of 
magnitude below this limit. Since we use an explicit method for the time integration, stability 
is an important issue. We find empirically that a time step smaller 
than $1/N$ leads to a stable time evolution. In 
Table~\ref{tablep} we give the parameters used in the numerical 
computations and the 
obtained numerical errors for different values of 
$\epsilon$. We note that the code is very efficient. Up to values of 
$\epsilon=10^{-3}$ the code can be run on standard computers without 
problems. 
\begin{table}[htb]
    \centering
    \begin{tabular}{|c|c|c|c|c|}
	\hline
	$-\log_{10}\epsilon$ & $\log_{2}N$ & $L$ & $\Delta t$ & $\log_{10}
       \mbox{err}$  \\
	\hline
	1 & 10 & 5 & $4*10^{-4}$ & -6.32 \\
	\hline
	1.25 & 12 & 5 & $2*10^{-4}$ & -7.79\\
	\hline
	1.5 & 12 & 5 & $2*10^{-4}$ & -6.33\\
	\hline
	1.75 & 14 & 5 & $10^{-4}$  &-6.30\\
	\hline
	2 & 14 & 5 & $5*10^{-5}$  &-6.29\\
	\hline
	 2.25 & 16 & 4  & $2.5*10^{-5}$ & -6.30\\
	 \hline
	  2.5 & 16 & 4  & $2.5*10^{-5}$ & -4.79\\
	  \hline
	  2.75 & 17 & 4 & $6.67*10^{-6}$  &-6.16\\
	  \hline
	  3 & 17 & 4 & $6.67*10^{-6}$  &-4.68\\
	\hline
    \end{tabular}
    \caption{Parameters used in the numerical integration of the KdV 
    equation for several values of $\epsilon$}    
    \label{tablep}
\end{table}

\section{Numerical solution of the Whitham  equations and of the Hopf 
equation}
In this section we solve numerically the Whitham equations (\ref{whitham}) 
for given initial data by
inverting  the hodograph transform (\ref{hodograph}) 
to obtain $\beta_1>\beta_2>\beta_3$ as a function of $x$ and $t$. 
Since the hodograph transform becomes degenerate at the leading and 
trailing edge 
we solve the system (\ref{lead0}) instead of (\ref{hodograph}) 
to avoid convergence problems.
In  a similar way we address the implicit solution of the Hopf equation 
(\ref{Hopfsol}). 

Both sets of equations are of the form 
\begin{equation}
    S_{i}(\{y_{i}\},x,t)=0,\quad i=1,\ldots,M
    \label{Si},
\end{equation}
where the $S_{i}$ denote some given real  function of the $y_{i}$ and $x$, 
$t$. The task is to determine the $y_{i}$ in dependence of $x$ and 
$t$. To this end we determine the $y_{i}$ for given $x$ and $t$ as 
the zeros of the function $S:=\sum_{i=1}^{M}S_{i}^{2}$. This will be 
done numerically by using the algorithm of \cite{optim} which is 
implemented as the function \emph{fminsearch} in Matlab. The 
algorithm provides an iterative approach which converges in our case 
rapidly if the starting values are close enough to the solution (see 
below how the starting values are chosen). Generally the iteration is 
stopped for function values smaller than $10^{-6}$. For quantitative 
comparisons we calculate the zeros to the order of machine precision. 

The solution $\beta_1(x,t)>\beta_2(x,t)>\beta_3(x,t)$ of the Whitham 
equations exists for $t>t_c$
where $t_c$ is the time of gradient catastrophe for the solution of the 
Hopf equation $u_t+6uu_x=0$. 
We look for a solution of the hodograph transform for 
$t>\dfrac{\sqrt{3}}{8}$ for a discretized time starting with time 
$t_{1}$ close to $t_{c}$.
For the moment we suppose that  $t<T$, 
where $T$ is defined in (\ref{T}).
For each fixed time, our strategy to solve (\ref{hodograph})
is the following:
First we solve the system (\ref{lead}) to obtain 
the leading edge coordinate
$x^{-}(t)$ and \[
\beta_1^{-}(t)>\beta^{-}_2(t)=\beta^{-}_3(t).\]
At  time $t_1$ 
the starting point for solving numerically (\ref{lead}) is given 
by $x^{-}_{0}\sim x_c$ and $\beta_i^0\sim u_c$. Similarly we solve the equations 
(\ref{trail}) for $x^{+}$ and $\beta_{1}=\beta_{2}$ and $\beta_{3}$ 
which fixes the interval $[x^{-},x^{+}]$. This interval is subdivided 
in a number of points $x_{n}$, $n=1,\ldots,N_{x}$. We use the values 
of the $\beta_{i}$ at point $x_{i-1}$ as starting values for the 
iteration at point $x_{i}$. A typical plot is 
shown in Fig.~\ref{fig6b}. Since the values of the 
$\beta_{i}$ change rapidly with $x$ near the leading and the 
trailing edge, we use a grid with one third of the points located in 
the vicinity of each of the edges and some wider grid spacing in 
between. To explore the parameter space in $t$ and $x$ we typically 
use $N_{x}=30$, for precision calculations we use $N_{x}=300$ and 
higher.  Additional points for plotting purposes 
are obtained via cubic interpolation.

For $t\geq T$, the procedure to solve the Whitham equations remains 
roughly  the same.
However we have to take care of the fact that 
the decreasing part of the initial data
contributes to the solution of the Whitham equations
when $\beta_3$ goes beyond the minimum value $-1$. Thus for $t>T$
we determine the value $x_{T}$ where $\beta_{3}=-1$ as a solution of 
(\ref{hodograph}). For $x>x_{T}$ we add the contributions of 
the decreasing part of the initial data. Since the 
algorithm \cite{optim} varies the values of the $\beta_{i}$ for fixed 
$x$, one has 
to make sure for values of $x$ near $x_{T}$ that the right branch of 
the logarithms is chosen. 

The functions (\ref{hodograph}) evaluated in the zero-finding 
algorithm contain elliptic integrals and 
integrals over the initial data which are calculated with a 
Chebychev collocation method. Elliptic integrals and functions are 
distributed with MATLAB where they are calculated with the 
algebro-geometric mean (see \cite{stegun}) to machine precision. 
We use here the approach to hyperelliptic functions via a Chebychev 
collocation method presented in \cite{cam}  (see also Chap.~6 of 
\cite{book}). The elliptic integrals and functions can be calculated 
to machine precision with this approach which is checked via internal 
tests and a comparison with the functions calculated with MATLAB. The 
reasons for the use of the Chebychev method are twofold. First we 
can apply similar techniques to calculate the non-standard integrals in 
(\ref{hodograph}) with machine precision. And secondly we develop in 
this way an approach to the one-phase Whitham equation which is directly open to a 
generalization to the multi-phase Whitham equations since the Chebychev code can 
handle hyperelliptic Riemann surfaces of in principle arbitrary genus. 

Let us briefly summarize the Chebychev approach, for details see 
\cite{cam,book,canuto,fornberg}. The Chebyshev
polynomials $T_n(x)$ are defined on the interval $I=[-1,1]$ by the
relation 
\[
T_n(\cos(t)) = \cos(n t)\;, \mbox{where } x = \cos(t)\;,
\qquad t\in[0,\pi]\;.
\]
The addition theorems for sine and cosine imply the recursion
relation
\begin{equation}
  \label{eq:recursderiv}
  \frac{T'_{n+1}(x)}{n+1} - \frac{T'_{n-1}(x)}{n-1} = 2 T_n(x)
\end{equation}
for their derivatives. The Chebyshev polynomials are orthogonal on $I$
with respect to the hermitian inner product
\[
\left< f, g \right> = \int_{-1}^1 f(x) \bar g(x) \,
\frac{\mathrm{d} x}{\sqrt{1-x^2}}\;.
\]
We have
\begin{equation}
  \label{eq:ortho}
  \left< T_m , T_n \right> = c_m \frac\pi2\, \delta_{mn}
\end{equation}
where $c_0=2$ and $c_l=1$ otherwise. A  function $f$ on $I$ is 
approximated via Chebychev polynomials, $f\approx 
\sum_{n=0}^{N}a_{n}T_{n}(x)$ where the spectral coefficients $a_{n}$
are obtained by the conditions $f(x_{l})=\sum_{n=0}^{N}a_{n}T_{n}(x_{l})$, 
$l=0,\ldots,N$. This approach is called a collocation method. If the 
collocation points are chosen to be 
$x_l=\cos(\pi l/N)$, the spectral coefficients follow from $f$ via  a 
Discrete Cosine Transform for which fast algorithms exist.

The fact that $f$ is approximated globally by a finite sum of
polynomials allows us to express any operation applied to $f$
approximately in terms of the coefficients. Let us illustrate this in
the case of integration. So we assume that $f = p_N =\sum_{n=0}^N a_n
T_n$ and we want to find an approximation of the integral for $p_N$,
i.e., the function
\[
F(x) = \int_{-1}^x f(s)\, \mathrm{d}s\;,
\]
so that $F'(x)=f(x)$. We make the ansatz $F(x) = \sum_{n=0}^N b_n\,
T_n(x)$ and obtain the equation
\[
F' = \sum_{n=0}^N b_n\,T'_n = \sum_{n=0}^N a_n T_n = f\;.
\]
Expressing $T_n$ in terms of the $T'_n$ via~(\ref{eq:recursderiv})
and comparing coefficients we get the equations
\[
b_1 = \frac{2a_{0} - a_{2}}{2}\;, \qquad b_n = \frac{a_{n-1} -
  a_{n+1}}{2n}\quad \mbox{for }0< n < N\;,\qquad b_N = 
  \frac{a_{N-1}}{2N}\;.
\]
between the coefficients which determine all $b_l$ in terms of the
$a_n$ except for $b_0$. This free constant is determined by the
requirement that $F(-1)=0$ which implies (because $T_n(-1)=(-1)^n$)
\[
b_0 = - \sum_{n=1}^N (-1)^n b_n\;.
\]
From the coefficients $b_n$ we can also find an approximation to the
definite integral $\int_{-1}^1 f(s)\,\mathrm{d}s = F(1)$ by evaluating 
\[
q_N(1) = \sum_{n=0}^Nb_n = 2\sum_{l=0}^{\lfloor N/2\rfloor}b_{2l+1}\;.
\]
Thus to find an approximation of the integral of a function $f$ we
proceed as described above, first computing the coefficients $a_n$ of
$f$, computing the $b_n$ and then calculating the sum of the odd
coefficients. 

There are two types of integrals in (\ref{hodograph}). Integrals of 
the form 
\begin{equation}
    \int_{a}^{b}\frac{f(\mu)d\mu}{\sqrt{\mu-a}}
    \label{whithint1}
\end{equation}
are calculated with the Chebychev integration routine after the 
transformation $\mu=a+(b-a)(1+y)^{2}/4$ where $y\in I$. The reason for 
this transformation is to obtain a smooth integrand free of square roots which 
is important for an efficient use of spectral methods. After the 
square root substitution we map the integration to the interval $I$ 
with a linear transformation. The second type of integral is of the 
form 
\begin{equation}
    \int_{-1}^{1}\frac{f(\nu)d\nu}{\sqrt{1-\nu^{2}}}
    \label{whithint2}.
\end{equation}
This integral is calculated by expanding the function $f$ in terms of 
Chebychev polynomials and by using the orthogonality relation 
(\ref{eq:ortho}). The precision of the numerical calculation is 
controlled via test integrals of known functions and via the spectral 
coefficients: the latter have to be of the order of machine precision 
for $n\sim N$ to provide sufficient resolution. In our examples 32 to 
128 polynomials were always sufficient to fulfill this requirement. 
Note that there are problems with the integrands if $\beta_{3}\sim-1$ 
since the $f_{\pm}$ diverge there. This would require an additional 
coordinate transformation to obtain a smooth integrand necessary for 
an efficient spectral approximation. Here we are, however, able to 
perform the required integration by hand. This has the additional 
benefit to provide a faster code, but the method is able to handle 
general initial data of the considered form.

\begin{figure}[!htb]
\centering
\epsfig{figure=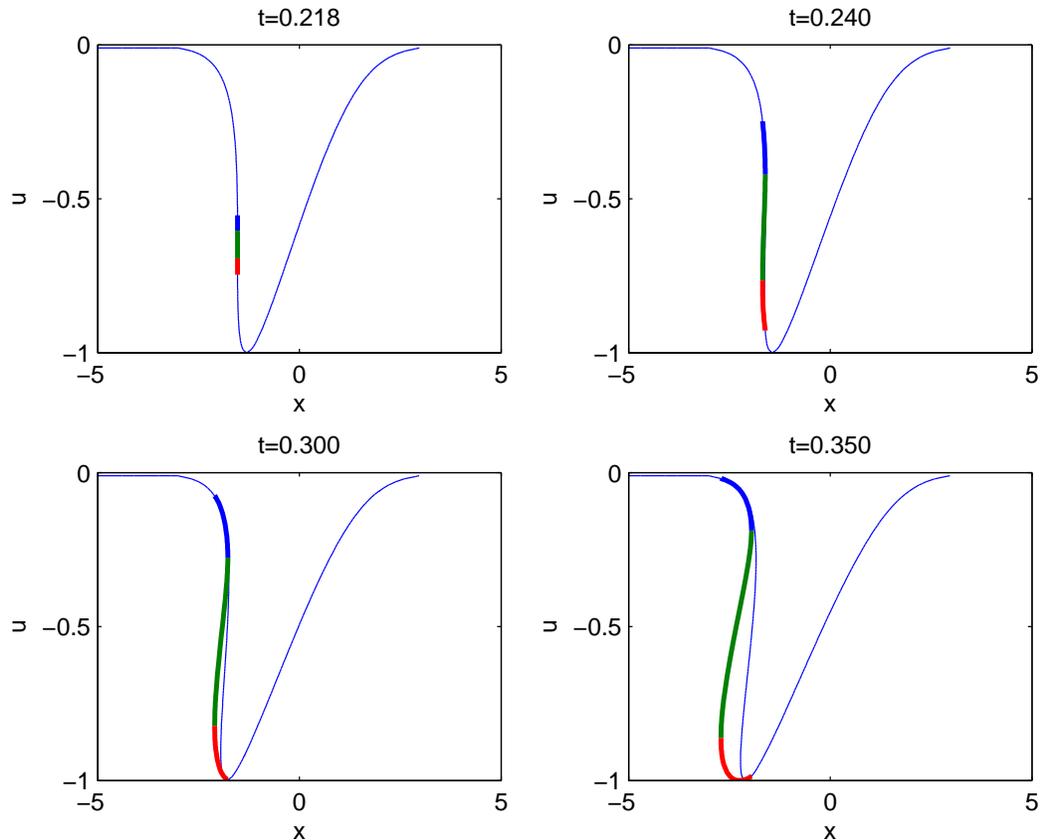, width=\textwidth}
\caption{Solution of the Hopf equation and the Whitham equation 
(thick line) 
for the initial data $u_0(x)=-1/\cosh^2x$ at various times.}
\label{figwhithopft}
\end{figure}

To obtain a solution of the Hopf equation  we choose again a convenient numerical 
grid in $t$ and $x\in[x_{L},x_{R}]$, where $x_L$ and $x_R$ are $x$-values  where the  solution 
is single valued.
We solve equation (\ref{Hopfsol}) for fixed 
$t$ and all corresponding values of $x$ starting from $x_{L}$ to $x^-$
with starting value $\xi=x$ as described above. Successively we 
repeat the procedure for greater values of $x_{i}$ on the grid with 
starting value $\xi_{i-1}$, the value of $\xi$ found in the preceding 
step. We find that this approach works well even for $x=x^{-}$  very 
close to the point of  gradient catastrophe $x_c$. Similarly we solve the Hopf 
equation starting from $x_{R}$ to 
$x^{+}$, the $x$-coordinate of the trailing edge. The solution of the Hopf equation in 
the region where it is multivalued, is obtained in 
Fig.~\ref{figwhithopft} by plotting the contour of zero values of the 
function $x-6tu\pm \ln ((1+\sqrt{1+u})/\sqrt{-u})$. It can be seen 
form Fig.~\ref{figwhithopft} that the matching of the solution to the Whitham equations
to the solution of the Hopf equation is  always $C^{1}$, but the solution 
of the latter 
in the multivalued region does not coincide with the former. 
The Whitham zone grows in time as can be seen from Fig.~\ref{figwhitt}.

To sum up we have shown in this section that we are able to obtain 
the numerical solution 
of the Whitham equations and the asymptotic approximation to the small 
dispersion limit of  the KdV equation with machine precision. The 
given method is in principle open to deal with general hump-like data 
with a single minimum. A generalization to a higher genus situation 
appears to be straight forward. 
Numerical solutions of the Whitham equations in the genus two case has
been obtained in \cite{JMS} for the Benjamin-One equation.

\section{Comparison of the small dispersion KdV solution and its  approximation}

\begin{figure}[!htb]
\centering
\mbox{\subfigure[KdV solution]
{\epsfig{figure=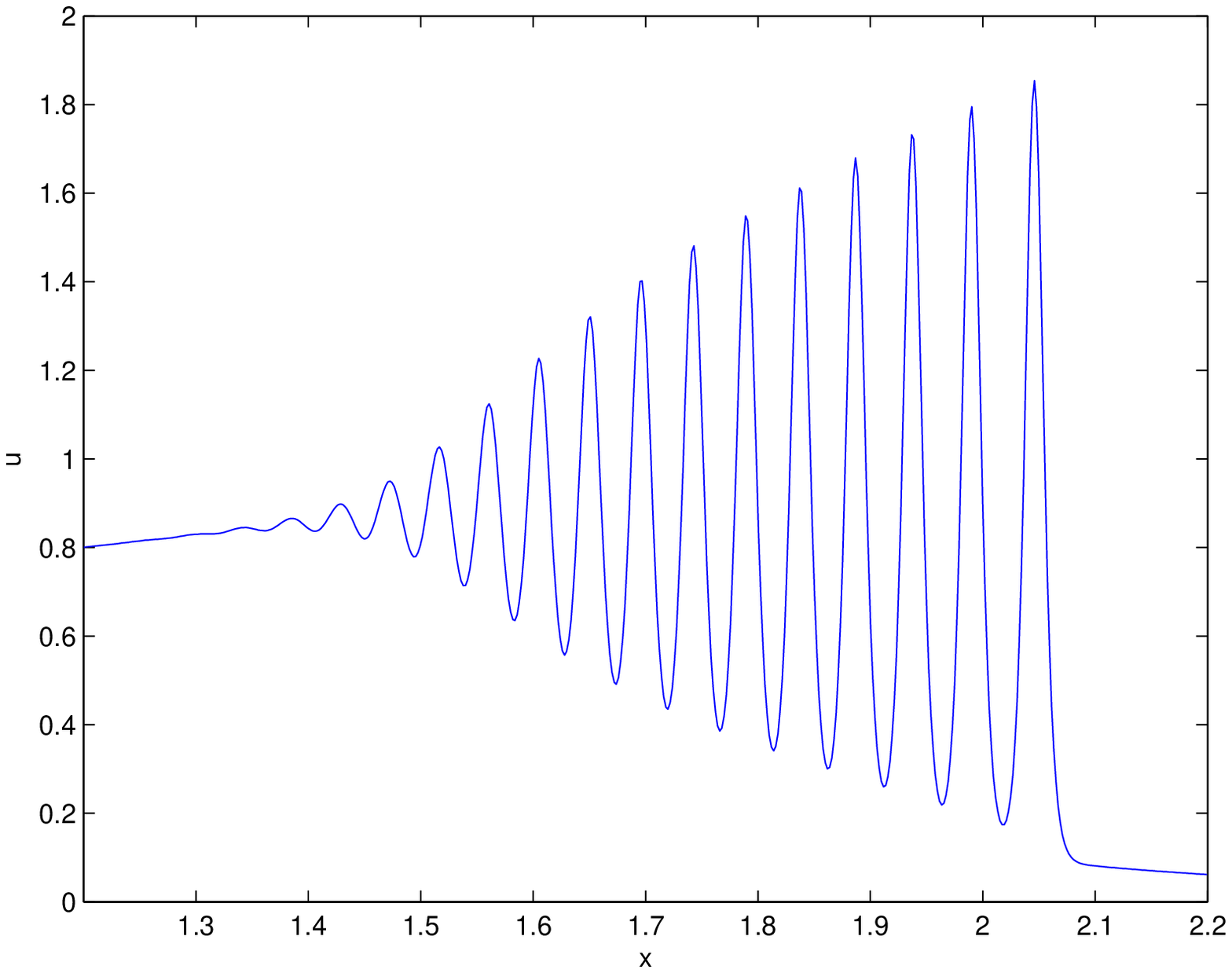, width=.5\textwidth}}\quad
\subfigure[Asymptotic solution]
{\epsfig{figure=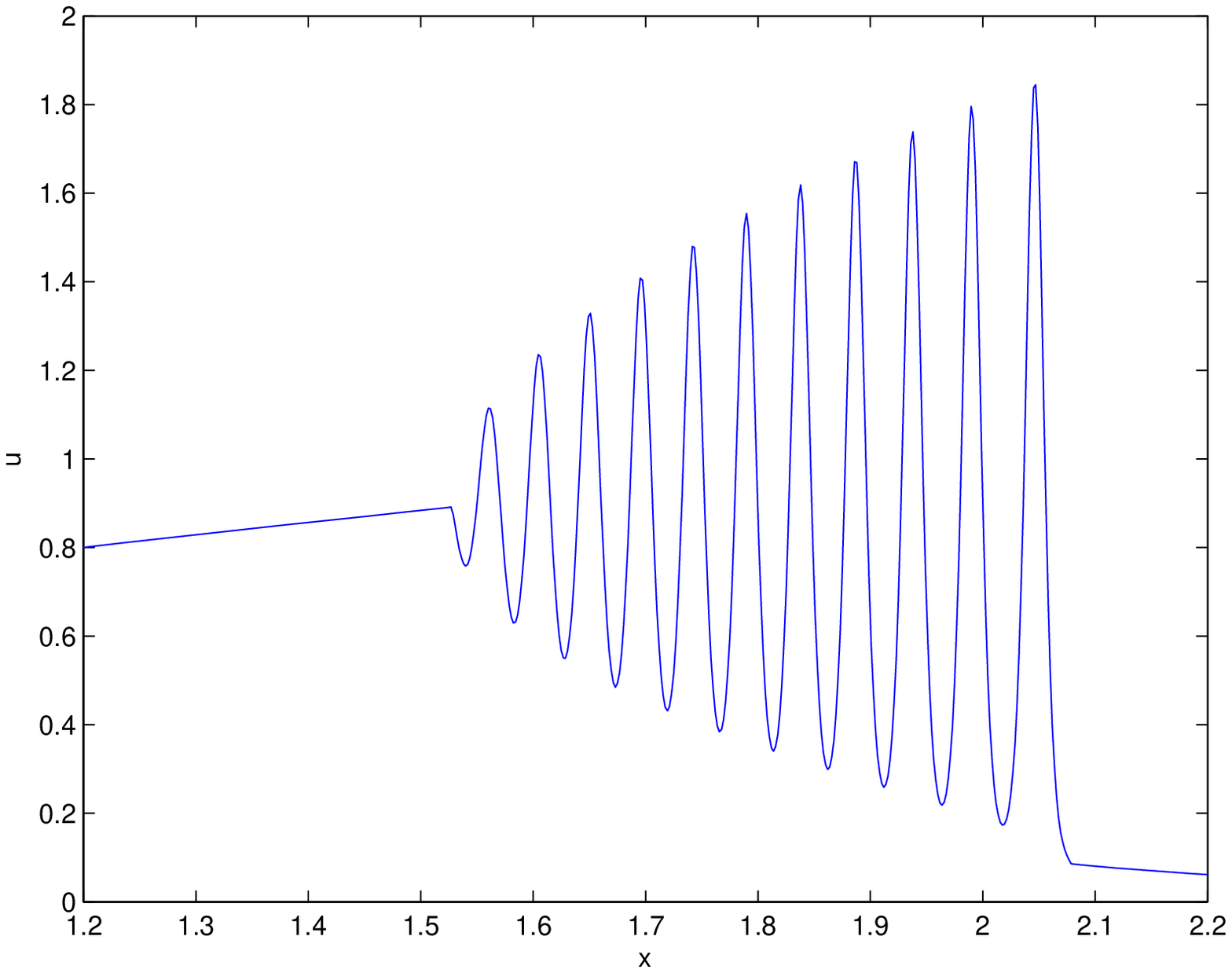, width=.5\textwidth}}}\newline
\caption{In  (a) the numerical solution of KdV  for the positive initial data
$u_0(x)=1/\cosh^2x$ and  in (b) the corresponding
 asymptotic formula (\ref{elliptic}) and 
(\ref{Hopfsol}) are plotted for   $t=0.35$ and  
$\epsilon=10^{-2}$.}
\label{figposdata}
\end{figure}

In the previous sections we have shown how to obtain numerical solutions to 
the KdV equation in the  small dispersion limit as well as the asymptotic solution (\ref{elliptic}) which 
follows from the solution of the  Whitham equations, both with controlled precision. 
This enables us to present a quantitative comparison 
of the KdV solution and the asymptotic approximation for several 
values of the dispersion parameter $\epsilon$. The code works for 
general rapidly decreasing initial data with a single hump, 
for positive initial data, see Fig.~\ref{figposdata}. This shows 
numerically that the formula for the phase in Theorem 2.2, which was 
originally obtained  only for negative initial data, can also be used for positive 
initial data.

\begin{figure}[!htb]
\centering
\epsfig{figure=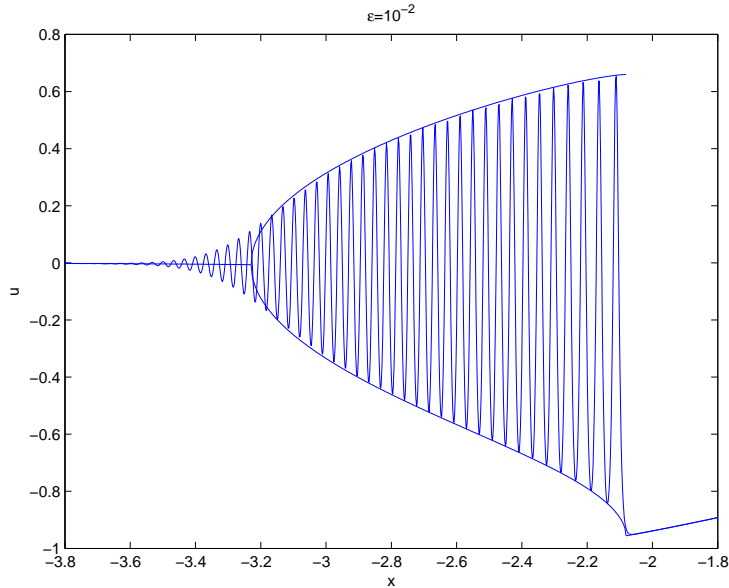, width=.7\textwidth}
\caption{Solution of the KdV equation with initial data 
$u_{0}=-1/\cosh(x)^{2}$ for $t=0.4$ together with the solution 
(\ref{Hopfsol}) and the envelope of the solution (\ref{elliptic}) for 
$\epsilon=0.01$.}
\label{figenvelope}
\end{figure}

In the following we will study as an example the initial data 
\[
u_0(x)=-\dfrac{1}{\cosh^2x}.
\]
The quality of the numerics allows to reach values of 
$\epsilon$ of the order of $10^{-3}$ without problems. For small 
$\epsilon$ the  asymptotic approximation is almost identical to 
the KdV solution in the Whitham zone as can be seen in 
Fig.~\ref{fig4}. Thus for small $\epsilon$ 
it makes little sense to plot both solutions 
in one figure as in Fig.~\ref{fig2in1} for $\epsilon=10^{-1}$. 
We show the envelope of the asymptotic solution 
(\ref{elliptic}) and the solution (\ref{Hopfsol}) together with the 
solution of the KdV equation in Fig.~\ref{figenvelope}. The shape of 
the envelope follows from (\ref{udn}) to be given by
$\beta_{1}-\beta_{2}+\beta_{3}$ and  $\beta_{1}+\beta_{2}-\beta_{3}$.

\begin{figure}[!htb]
\centering
\epsfig{figure=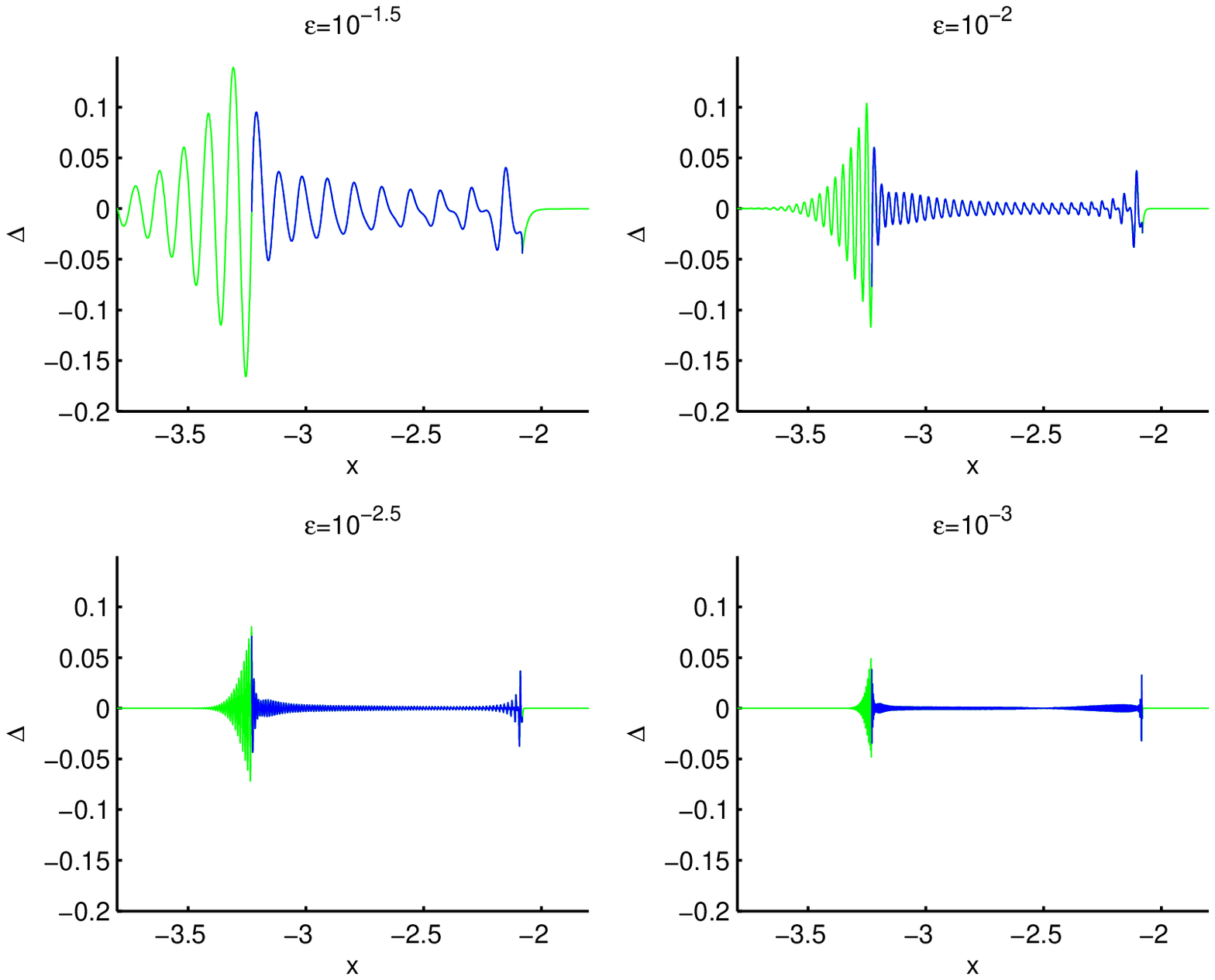, width=\textwidth}
\caption{The blue line describes the  difference between the 
numerical solution of the KdV equation  and
the asymptotic formula (\ref{elliptic})
for the initial data $u_0(x)=-1/\cosh^2x$ and for $t=0.4$.
The green lines represent the  difference between the 
numerical solution of the KdV equation  and the Hopf solution 
(\ref{Hopfsol}).}
\label{figc2}
\end{figure}

The numerical precision enables us to study quantitatively the 
difference between the KdV solution and the asymptotic solutions  
(\ref{elliptic}) and (\ref{Hopfsol})  and to establish where the 
approximation is satisfactory and where it is not. We show this difference 
for various values of $\epsilon$ in Fig.~\ref{figc2} at time 
$t=0.4$. 

From Fig.~\ref{figc2} and Fig.~\ref{fig5}, it is clear that the error is 
decreasing with $\e$. Indeed in the
first plot, the highest peak is of the order of $0.15$, while in the last plot 
it is of the order $0.06$. It is also obvious that the asymptotic formula 
(\ref{elliptic})  gives a 
satisfactory description of the oscillations within an interval in the Whitham region
$[x^-(t),x^+(t)]$. At the boundaries of this zone, the highest peaks 
in the difference appear. 

A similar behavior can be observed for the asymptotic solution 
(\ref{Hopfsol}). The smaller $\e$, the better the approximation, 
which is always worst at the boundary of the Whitham 
zone. It is also clear that there are in any case oscillations of the 
KdV solution for values of $x<x^{-}(t)$ for $t>t_{c}$ whereas the 
solution (\ref{Hopfsol}) of the Hopf equation, which is considered as 
an approximation there, obviously does not show any oscillations. As 
can be seen from Fig.~\ref{figc2}, this oscillatory zone shrinks when 
$\epsilon$ becomes smaller. For values of $x>x^{+}(t)$, no 
oscillations are observed but  the difference between the KdV solution 
and the solution (\ref{Hopfsol}) is biggest at the boundary of the 
Whitham zone and goes  asymptotically to zero. The zone where the 
solutions differ considerably also shrinks with decreasing 
$\epsilon$. 
Below we will study these features in more detail.

We will present certain characteristic quantities as the difference 
between the exact and the approximative solution in dependence of 
$\epsilon$ and show loglog-plots of these quantities. If 
appropriate we perform a linear regression analysis to identify a 
scaling behavior in $\epsilon$ of the studied quantity. We briefly 
summarize the basic relations for a linear regression (see e.g.\ 
\cite{acton}). Given a set of real points $y_{i}$, $z_{i}$ with 
$i=1,..,M$, we perform a least square fitting to the line $y=az+b$ 
where with ($\bar{z}$, $\bar{y}$ are the mean values)
\begin{equation}
    s_{zz}=\sum_{i=1}^{M}(z_{i}-\bar{z})^{2},\quad s_{zy}
    =\sum_{i=1}^{M}(z_{i}-\bar{z})(y_{i}-\bar{y}),\quad 
    s_{yy}=\sum_{i=1}^{M}(y_{i}-\bar{y})^{2}
    \label{sxx},
\end{equation}
the parameters $a$ and $b$ are given by
\begin{equation}
    b = \frac{s_{zy}}{s_{zz}},\quad a = \bar{y}-b\bar{z}
    \label{ab}.
\end{equation}
For the correlation coefficient being a measure of the quality of 
the fitting, and the for the standard errors of $a$ and $b$ one has
\begin{equation}
    r = \frac{s_{zy}}{\sqrt{s_{zz}s_{yy}}},\quad 
    \sigma=\sqrt{\frac{s_{yy}-bs_{zy}}{M-2}},\quad 
    \sigma_{a}=\sigma\sqrt{\frac{1}{M}+\frac{\bar{z}^{2}}{s_{zz}}},\quad
    \sigma_{b}=\frac{\sigma}{\sqrt{s_{zz}}}
    \label{rsigma}.
\end{equation}

We will only present the results of linear regression if the 
correlation coefficient is at least of the order of $0.99$. Even in 
these cases the results have to be taken with care. We only use 
a small number of points, but more importantly, we only study 
values of $\epsilon$ between $10^{-1}$ and $10^{-3}$. Thus these 
plots show  a scaling law in this regime. Further analytical work 
will have to show whether these results hold for more general values 
of $\epsilon$, and what are the precise values of the coefficients.

\begin{figure}[!htb]
\centering
\epsfig{figure=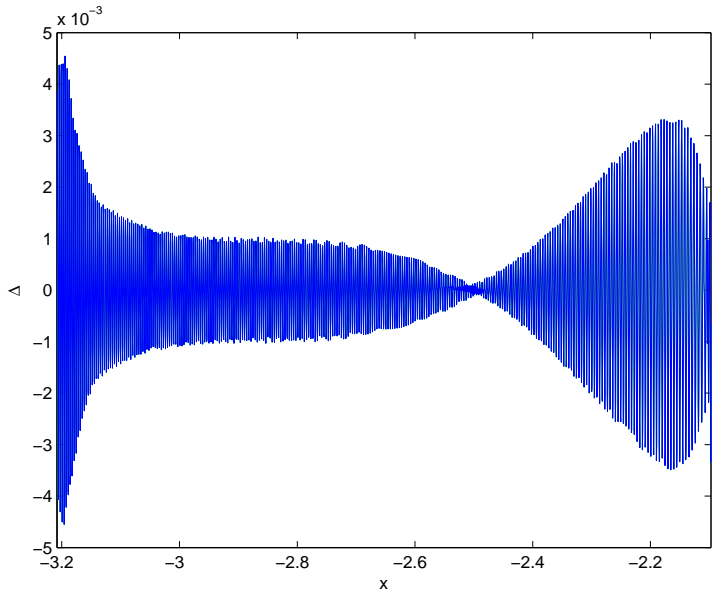, width=.7\textwidth}
\caption{Difference between the 
numerical solution of the KdV equation  and
the asymptotic formulas  (\ref{elliptic}) for 
$t=0.4$ and $\epsilon=10^{-3}$ in the `interior' Whitham zone.}
\label{figwhithint}
\end{figure}

\paragraph{\it Whitham zone}
From Fig.~\ref{figc2} and Fig.~\ref{fig5} it is obvious that the 
asymptotic approximation (\ref{elliptic}) does not give a 
satisfactory description of the KdV solution close to 
the boundary. However one can identify an interior zone where this 
approximation gives a rather accurate description of the KdV zone. 
There is an obvious arbitrariness in the definition of such an 
interior zone since the difference between KdV solution and 
approximation is not constant there. As can be seen from 
Fig.~\ref{figwhithint}
there is also an error in the phase. Close to $x=-2.5$ the KdV 
solution and the asymptotic solution are in phase, and consequently 
the difference is minimal there. 

A possible definition of the interior zone is simply to cut off the big 
oscillations at  the edges. If one cuts off just one full 
oscillation, the cut off zone scales roughly like $\epsilon$ in 
accordance with the fact that there are
oscillations  of the order  $1/\epsilon$  in the Whitham zone.
For the case $t=0.4$ and $\epsilon=10^{-3}$, 
we obtain Fig.~\ref{figwhithint}.

\begin{figure}[!htb]
\centering
\epsfig{figure=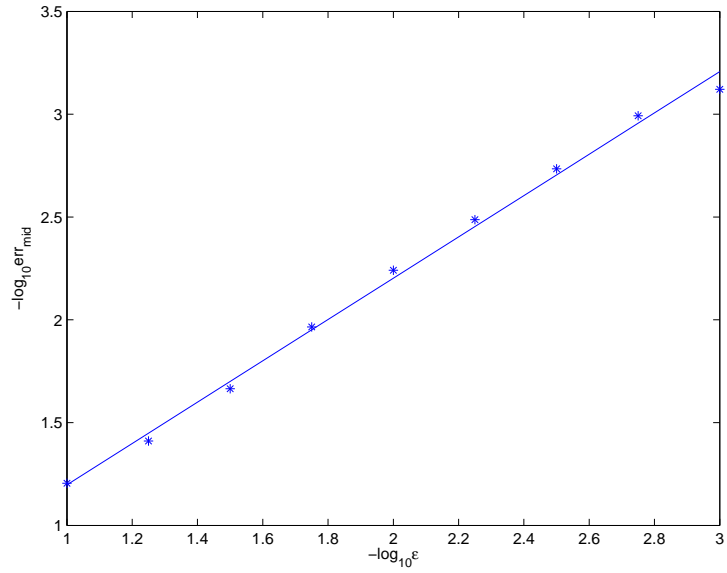, width=.7\textwidth}
\caption{Maximum of the difference  between the KdV
numerical solution   and
the asymptotic formula (\ref{elliptic}) in the interior Whitham zone for 
$t=0.4$ in dependence of $\epsilon$; the data can be fitted by a 
straight line 
$y=az+b$ with $a=1.0049$ and 
    $b=0.1005$.}
\label{figerrorw}
\end{figure}

To define an error  ($\mbox{err}_{mid}$) in the interior
we take the maximum 
of the absolute value of the difference between the solutions in the 
vicinity of 
the center of the zone ($x\approx -2.66$). Notice that there is a 
certain crudeness in  this definition of the error since this maximum 
will occur for different $\epsilon$ at slightly different values of 
$x$. Because of the error in the phase there are large differences in the 
error for different $\epsilon$ if one takes the difference at the 
same $x$ value in all cases. The so defined error is shown in 
Fig.~\ref{figerrorw}, where it can be seen that it decreases almost
linearly with $\epsilon$. Linear regression analysis yields $a=1.0049$,
$b=0.1005$ with $\sigma_{a}=0.05$, $\sigma_{b}=0.024$, and a 
correlation coefficient $r=0.998$.  

\begin{figure}[!htb]
\centering
\epsfig{figure=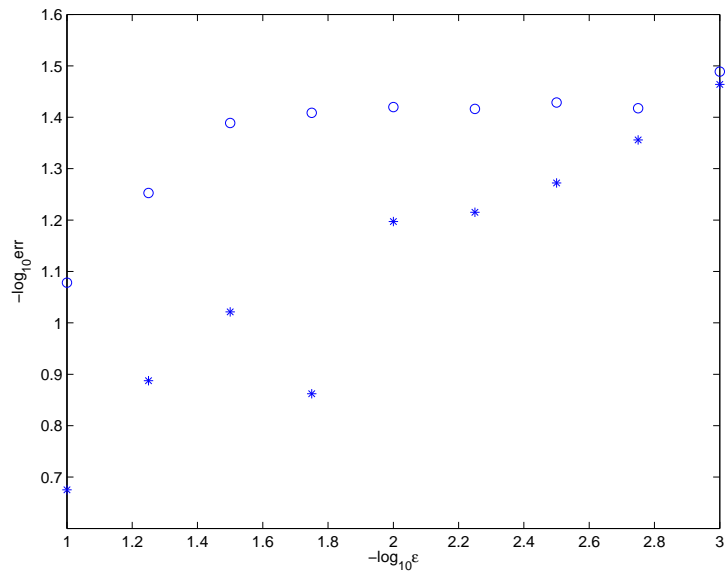, width=.7\textwidth}
\caption{Maximum of the difference  between the KdV
numerical solution   and
the asymptotic formula (\ref{elliptic}) at the boundaries of Whitham 
zone for $t=0.4$ in dependence of $\epsilon$ ($*$ near $x^{-}$ and 
$o$ near $x^{+}$).}
\label{figerrormaxwh}
\end{figure}

The maximal error near the boundaries of the Whitham zone is shown in 
Fig.~\ref{figerrormaxwh}. There seems to be no obvious scaling of 
these errors with $\epsilon$ which is probably due to the rapid 
oscillations of the error because of the error in the phase.

\begin{figure}[!htb]
\nonumber
\centering
\epsfig{figure=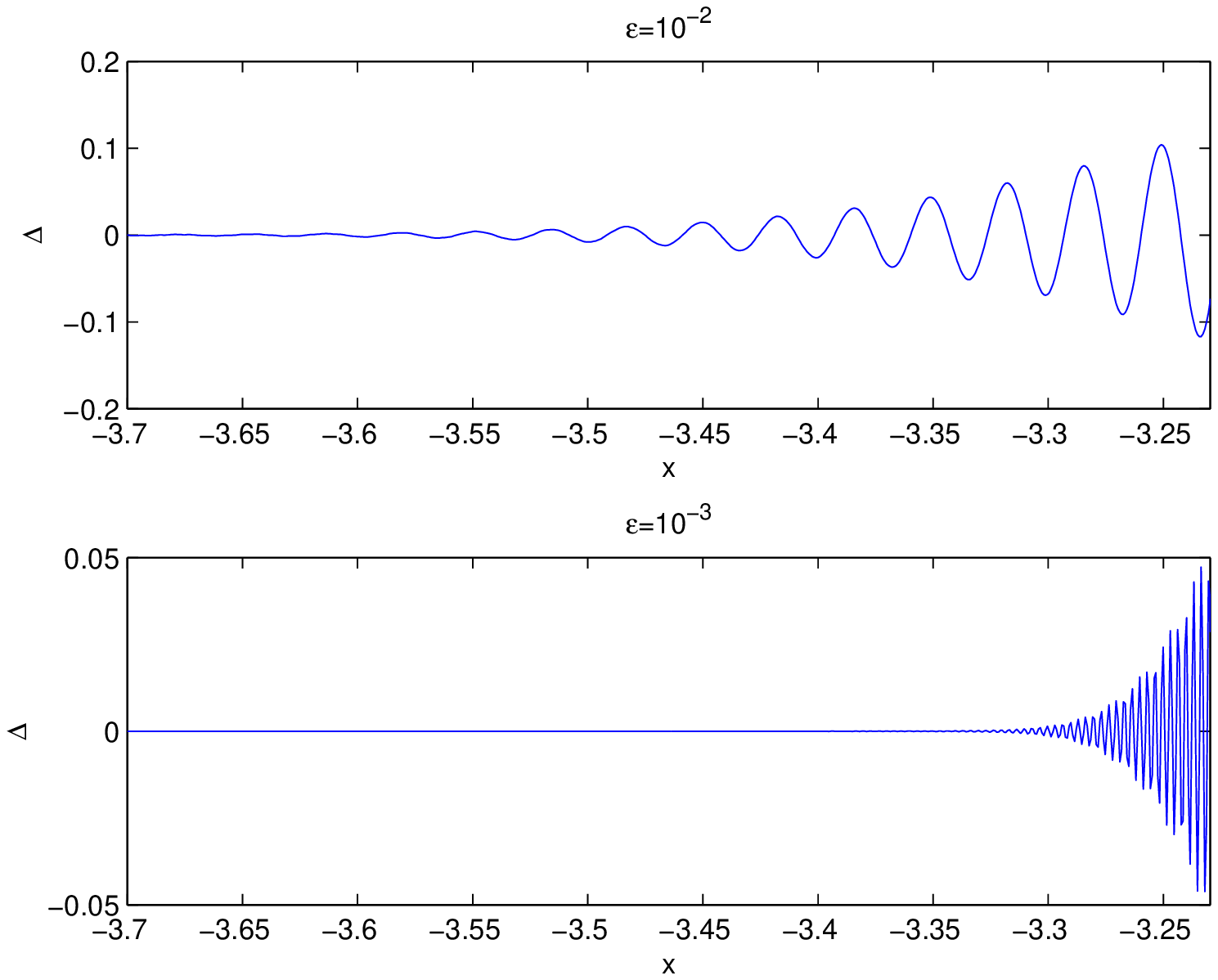, width=\textwidth}
\caption{Difference between the 
numerical solution of the KdV equation
 and the asymptotic formula 
(\ref{Hopfsol}) for $t=0.4$ and $x<x^-(t)$.}
\label{figleft}
\end{figure}

\paragraph{\it Oscillatory zone with $x<x^{-}(t)$}
As can be seen in Fig.~\ref{fig4} and 
Fig.~\ref{fig5}, there are always oscillations of the KdV 
solution for $t>t_{c}$ which do not occur for the asymptotic solution 
(\ref{Hopfsol}). We show this region in detail in Fig.~\ref{figleft}.

\begin{figure}[htb]
\centering
\epsfig{figure=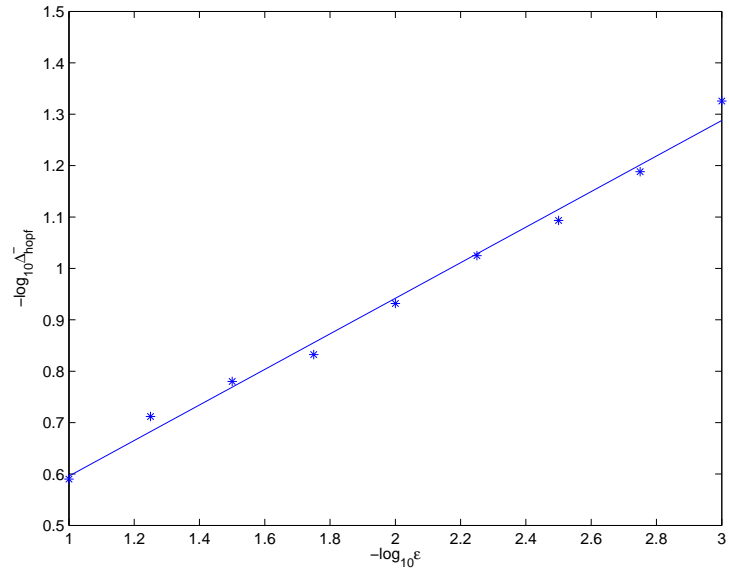, width=.7\textwidth}
\caption{The size of the left oscillatory zone $\Delta^-_{hopf}$ as a 
function of $\epsilon$ for $t=0.4$ in dependence of $\epsilon$; 
the data can be fitted by a straight line 
$y=az+b$ with $a=0.761$ and 
    $b=-0.789$.}
\label{figoscx}
\end{figure}

\begin{figure}[!htb]
\centering
\epsfig{figure=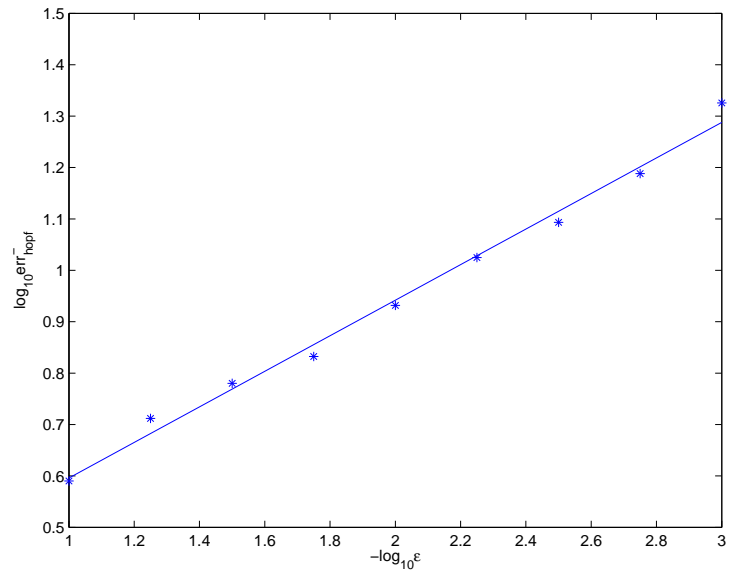, width=.7\textwidth}
\caption{Maximum of the absolute value of the difference 
$\mbox{err}_{hopf}^{-}$ between the 
numerical solution of the KdV equation  and
the Hopf solution  (\ref{Hopfsol}) for 
$t=0.4$ in dependence of $\epsilon$; the data can be fitted by a 
straight line 
$y=az+b$ with $a=0.346$ and 
    $b=0.25$.}
\label{figoscerror}
\end{figure}

It can be seen that the oscillatory zone clearly shrinks with 
decreasing $\epsilon$. There is no obvious rigorous definition of the 
end of the oscillatory zone. The difference will eventually be of the 
order of the numerical error for the KdV solution. We define the end 
of the oscillatory zone as the $x$-value
where the amplitude of the difference of the 
solutions is smaller than $10^{-4}$ which is of the order of the 
numerical accuracy we have used. The width of this zone, 
$\Delta_{hopf}^{-}:=x_{hopf}^{-}/x^{-}-1$ 
in dependence of $\epsilon$ is shown in Fig.~\ref{figoscx}. It can 
be seen that the oscillatory zone shrinks to zero  with $\epsilon$. 
Asymptotically the boundary of this zone approaches the boundary of 
the Whitham zone. We find that $\Delta_{hopf}^{-}$  scales roughly 
as $\epsilon^{3/4}$. The linear regression analysis yields $a=0.761$ 
and $b=-0.789$ with $\sigma_{a}=0.028$, $\sigma_{b}=0.013$ and 
$r=0.999$. However, this result has to be taken with care 
in view of the low number of points and the arbitrariness in the 
definition of the zone. 
The error  $\mbox{err}_{hopf}^{-}
$  in this zone is  measured by the maximum of the absolute 
value of the difference of the  amplitude between the KdV solution and the Hopf solution.
This maximum value always occurs close to the boundary of
the Whitham zone. The  error $\mbox{err}_{hopf}^{-}
$ is shown in Fig.~\ref{figoscerror}. It 
decreases roughly as $\epsilon^{1/3}$. Linear regression analysis 
yields $a=0.346$, $b=0.250$ with $\sigma_{a}=0.025$, 
$\sigma_{b}=0.012$ and $r=0.996$. 

\paragraph{\it Zone with $x>x^{+}(t)$}
A similar behavior is found in the zone for $x>x^{+}(t)$ with the 
exception that there are no oscillations. The difference of the 
asymptotic solution (\ref{Hopfsol}) and the KdV solution is biggest 
for $x^{+}(t)$ and decreases monotonically to the order of the numerical 
precision. 
\begin{figure}[!htb]
\nonumber
\centering
\epsfig{figure=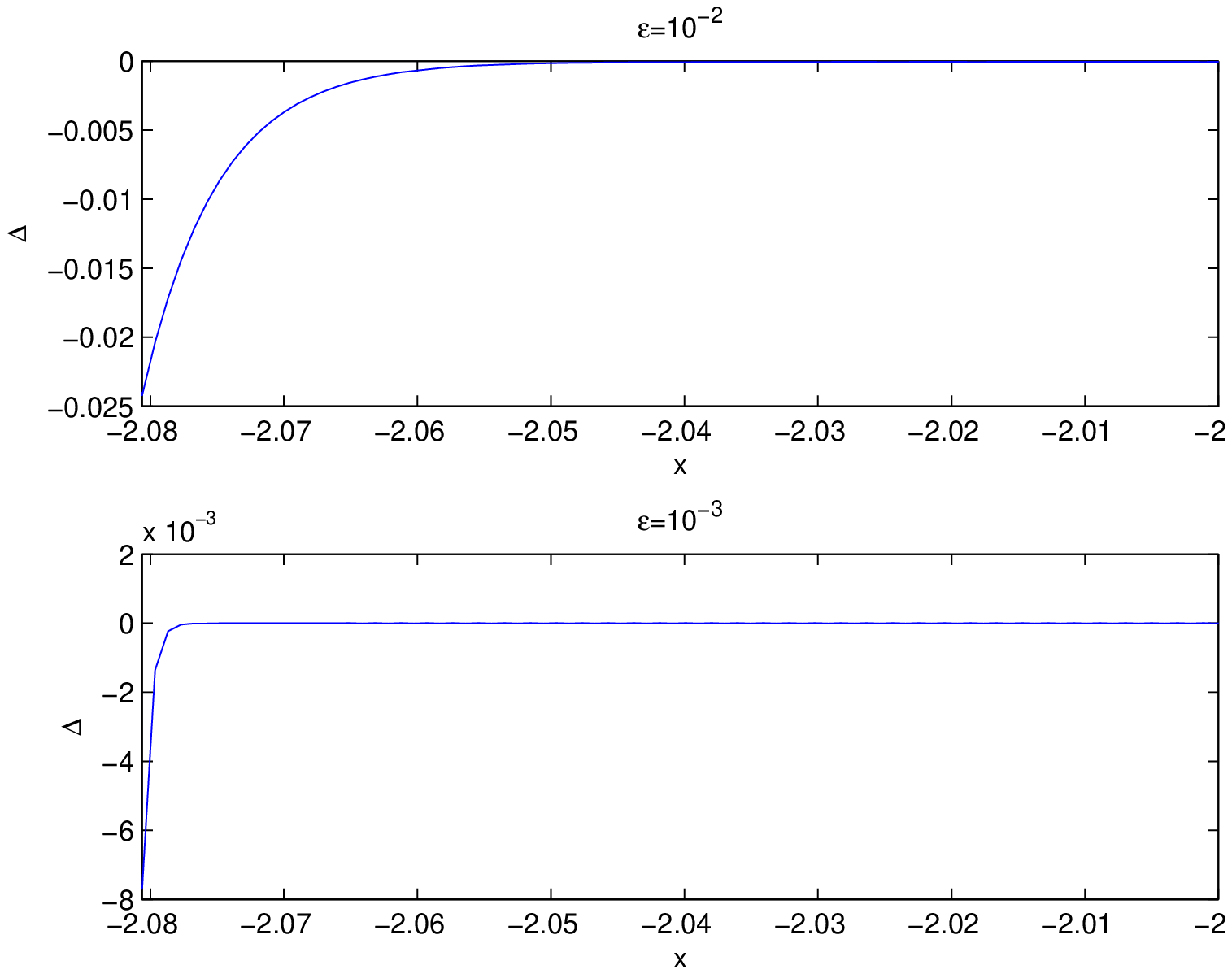, width=\textwidth}
\caption{Difference between the 
numerical solution of the KdV equation
 and the Hopf solution 
(\ref{Hopfsol}) for $t=0.4$ and $x>x^+(t)$.}
\label{figright}
\end{figure}

\begin{figure}[htb]
\centering
\epsfig{figure=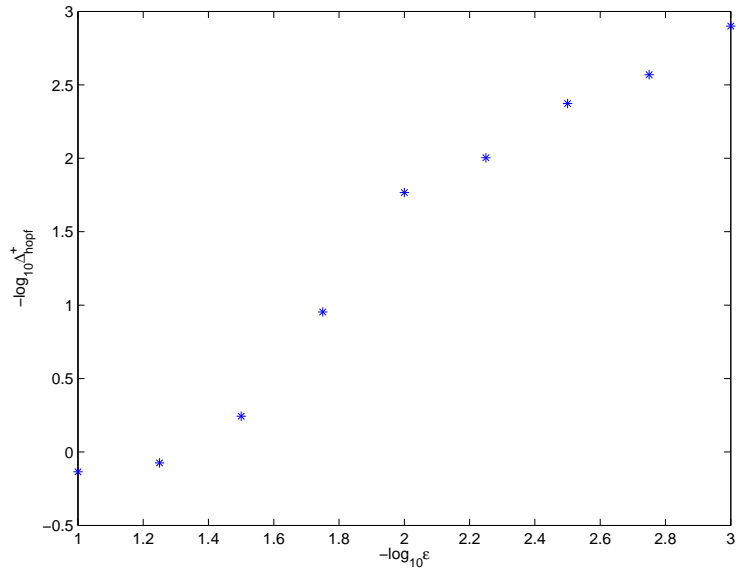, width=.7\textwidth}
\caption{The size of the zone $\Delta^+_{hopf}$ as a 
function of $\epsilon$ for $t=0.4$.}
\label{figdhopf}
\end{figure}

We define the width of the zone as the region where the absolute 
value of the difference is bigger than $10^{-4}$. The values of the 
quantity  $\Delta_{hopf}^{+}:=1-x_{hopf}^{+}/x^{+}$ 
are shown in Fig.~\ref{figdhopf} in dependence on 
$\epsilon$. It can be seen that this zone shrinks with $\epsilon$ to 
zero, i.e., its boundary approaches asymptotically the boundary of 
the Whitham zone. Apparently there is no simple $\epsilon$-dependence 
of this quantity, at least not for the number of points used in the 
plot.

\begin{figure}[htb]
\centering
\epsfig{figure=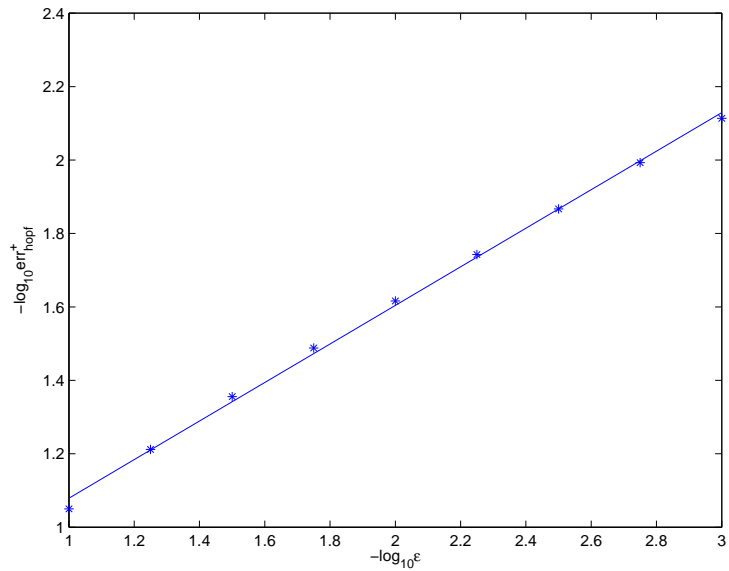, width=.7\textwidth}
\caption{The maximal error in the zone $x>x^{+}$ as a 
function of $\epsilon$ for $t=0.4$; the data can be fitted by a 
straight line 
$y=az+b$ with $a=0.525$ and 
    $b=0.554$.}
\label{figdeltamaxr}
\end{figure}

The maximal error in this zone always occurs at the boundary to the 
Whitham zone. As can be seen in Fig.~\ref{figdeltamaxr}, this error 
decreases roughly as $\sqrt{\epsilon}$. Linear regression analysis 
yields $a=0.525$, $b=0.554$ with $\sigma_{a}=0.017$, 
$\sigma_{b}=0.008$ and $r=0.999$. 

\begin{figure}[!htb]
\centering
\epsfig{figure=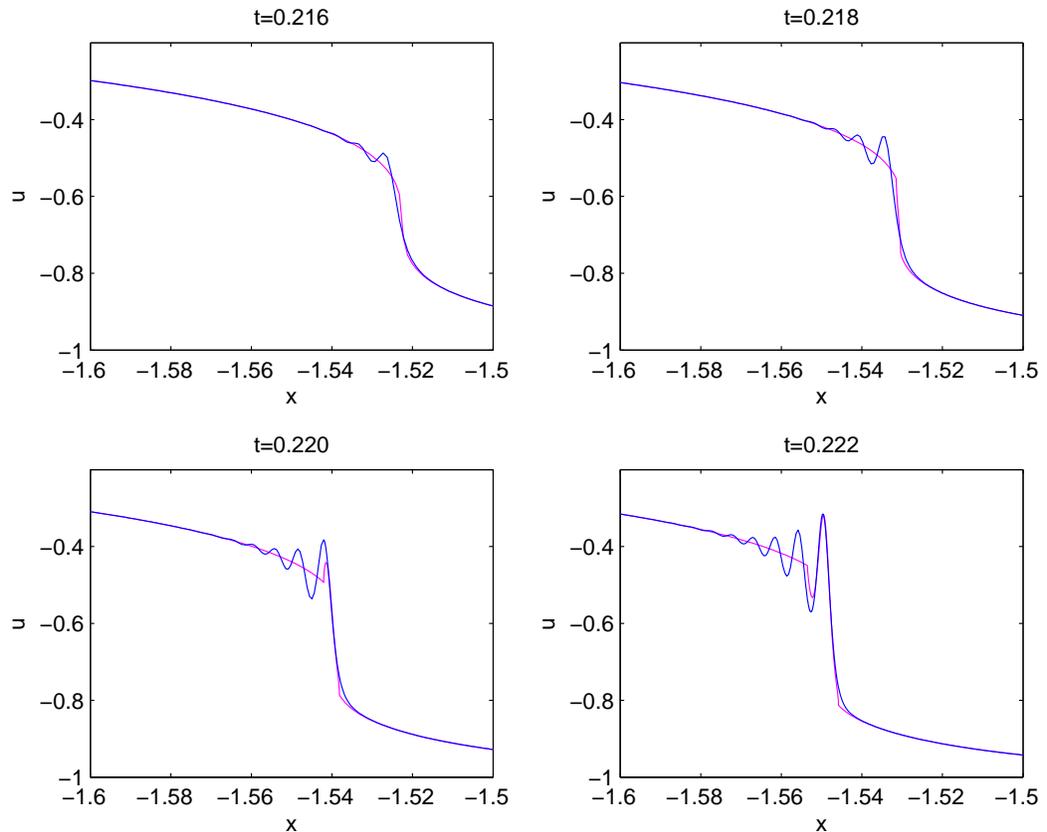, width=\textwidth}
\caption{KdV solution and asymptotic solution for $\epsilon=10^{-3}$ 
close to the breakup time.}
\label{figbreak4}
\end{figure}

\begin{figure}[!htb]
\centering
\epsfig{figure=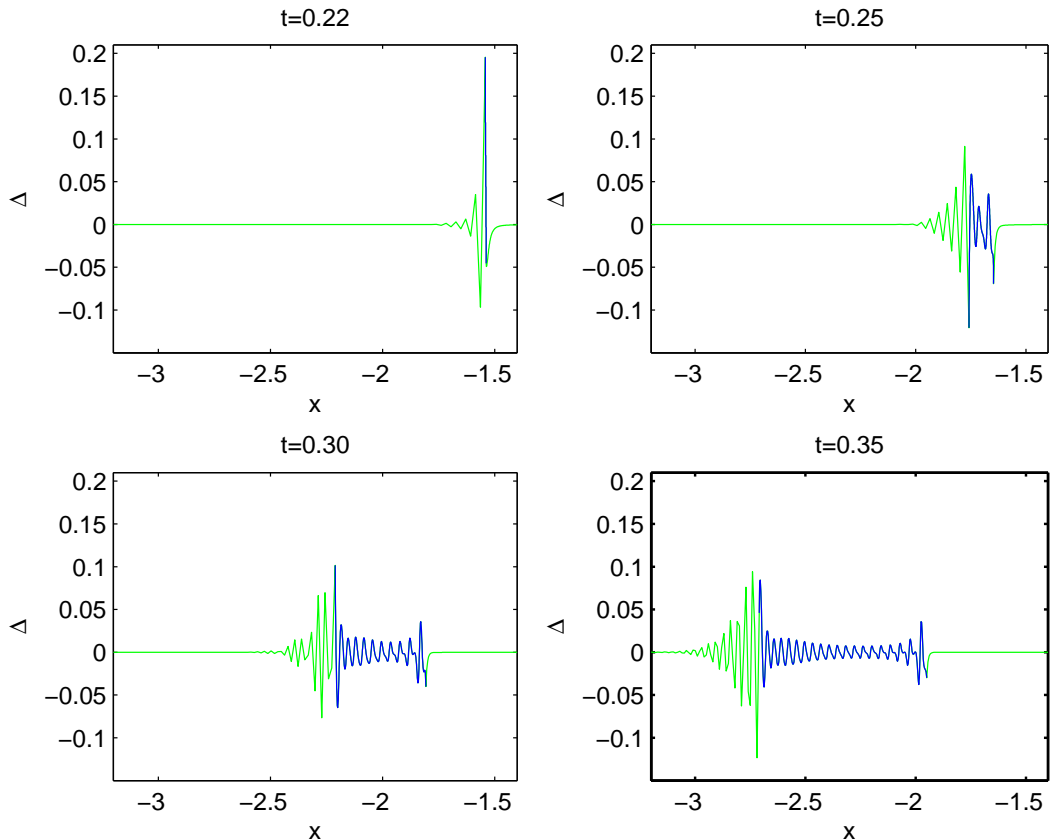, width=\textwidth}
\caption{Difference of the KdV solution and the asymptotic solution 
for $\epsilon=10^{-2}$ 
for different times.}
\label{figdifft4}
\end{figure}

\paragraph{\it Breakup time}
As discussed above the asymptotic approximation of the small 
dispersion KdV solution is always worst near the boundary of the 
Whitham zone. At the breakup time $t_{c}$ being defined as the time 
of gradient catastrophe of the solution to the Hopf equation,  
the Whitham zone consists only of one point. In Fig.~\ref{fig2} 
it can be seen that the solution to the KdV equation always forms 
oscillation for times $t<t_{c}$. Thus the first oscillation in the 
Whitham zone gives only a very crude approximation to the oscillation of 
the KdV solution in this zone. In addition there are typically 
several oscillation of the KdV solution outside the Whitham zone at 
this time. The discrepancy between the asymptotic and the KdV 
solution is quite pronounced even for small $\epsilon$ as can be 
seen from Fig.~\ref{figbreak4} for $\epsilon=10^{-3}$. The KdV 
solution develops roughly the same number of oscillations before 
$t_{c}$ as in Fig.~\ref{fig2}, but on a smaller scale. As already 
stated, the oscillatory zone shrinks with decreasing $\epsilon$, but 
close to the breakup the difference between 
the asymptotic solutions and the KdV solution remains considerable, 
even for small $\epsilon$. As can be seen in Fig.~\ref{figdifft4}, 
the difference between asymptotic solution and KdV solution is 
biggest close to the breakup time. 

\paragraph{\it Time dependence}
The qualitative behavior of the difference of the small dispersion 
KdV solution and the asymptotic solution as outlined above is 
characteristic for all times as can be seen from Fig.~\ref{figdifft4}. 
The difference is always biggest at the boundary of the Whitham zone. 
The absolute value of this difference is almost constant in time at 
the boundary. Therefore the discrepancy between the approximation and 
the solution is biggest close to breakup. The interior zone 
where the asymptotic solution gives a very satisfactory approximation 
grows with time. 

\section{Outlook}
In this paper we have presented a quantitative numerical treatment of 
the solution to the KdV equation in the small dispersion limit for 
initial data with compact support and a single hump. The same has been 
achieved for the asymptotic formulas for the same  initial data.
 The code for the KdV solution is set up to handle general 
initial data with compact support, thus the inclusion of initial data which develop multi-phase solutions
is  directly possible. The approach to 
the solution of the one-phase Whitham equations is also open to a generalization to 
multi-phase case. In particular the existence and uniqueness of the solution of the two-phase Whitham equations has been obtained in \cite{GT} for generic initial data where $f_-$ has negative fifth derivative. 
 Since the Chebychev code \cite{cam,book} to 
calculate the characteristic quantities on the elliptic surface in the 
present paper is able to deal with almost arbitrary hyperelliptic 
surfaces, the corresponding models can be studied. This will be the 
subject of future work.

The results of the previous section show that the asymptotic 
approximation is worst at break up point  and at the boundary of the Whitham 
zone. According to the conjecture  in \cite{D0},\cite{KS}
 the solution of the KdV equation, near the point of gradient catastrophe, 
is well described by 
\begin{equation}
\label{pain0}
u(x,t,\e)= u_c+  \left(\dfrac{\e}{  k}\right)^{2/7} F\left(- \dfrac{1}{\epsilon}\left(\dfrac{\e}{k}\right)^{1/7} 
( x - x_c - 6 u_c (t-t_c));\;\dfrac{1}{\e}\left(\dfrac{\e}{k}\right)^{3/7}(t-t_c)\right),
\end{equation}
where
\[
\quad k=-\dfrac{f_-'''(u_c)}{6},
\]
and the function $F(X;T)$ is the solution of the fourth order ODE of the  Painlev\'e I hierarchy
 \begin{equation}
\label{pain1}
-6 T F + F^3 + F F'' +1/2 (F')^2 +1/10 F'''' = X.
\end{equation}
The above equation first appeared in the double scaling limits of one 
matrix models for the multicritical 
index $m=3$ and for $T=0$ \cite{BMP}. It is conjectured that the equation (\ref{pain1}) has a smooth real 
solution on the real line with asymptotic behavior $F(X,0)=\pm X^{\frac{1}{3}}$ for $X\ra \pm\infty$.  
We will investigate numerically the asymptotic description given by (\ref{pain0}) in a subsequent publication.

The description of the left oscillatory zone outside the Whitham zone, should be obtained in the spirit
of matrix models, performing a sort of double scaling limit.
According to the ansatz in \cite{KS}, the envelope of the oscillations is given by 
the Hastings McLeod solution of  the second Painlev\'e equation. This result is in accordance with 
the rigorous results obtained in the double scaling
limit in one-matrix models \cite{DKMVZ},\cite{BI},\cite{CKV}.

Regarding the right boundary of the Whitham zone, at the moment, 
there is no theoretical ansatz for an 
asymptotic description. Our numerical results could possibly provide a hint for 
an ansatz for the asymptotic behavior.

It would also be very  interesting to perform similar numerical investigations for the semiclassical 
limits of the focusing \cite{KMM} and defocusing \cite{JLM} nonlinear Schr\"odinger equation.

\end{document}